\documentclass{aa}  
\usepackage[english]{babel}
\usepackage{graphicx}
\usepackage{txfonts}
\newcommand\Zmax{$Z_\mathrm{max}$}
\newcommand\z{$Z$}
\newcommand\R{$R$}

\newcommand\Rg{$R_\mathrm{g}$}
\newcommand\temp{$T_{\fontsize{6}{6}\selectfont \mbox{eff}}$}\normalfont
\newcommand\logg{$\log g$}
\newcommand\met{[M/H]}

\newcommand\RCa{$Z_\mathrm{0.0}^\mathrm{RAVE\_RC}$}
\newcommand\RCb{$Z_\mathrm{0.4}^\mathrm{RAVE\_RC}$}
\newcommand\RCc{$Z_\mathrm{0.8}^\mathrm{RAVE\_RC}$}
\newcommand\RCd{$Z_\mathrm{1.2}^\mathrm{RAVE\_RC}$}

\newcommand\RCma{$Z_\mathrm{0.0}^\mathrm{mock\_RC}$}
\newcommand\RCmb{$Z_\mathrm{0.4}^\mathrm{mock\_RC}$}
\newcommand\RCmc{$Z_\mathrm{0.8}^\mathrm{mock\_RC}$}
\newcommand\RCmd{$Z_\mathrm{1.2}^\mathrm{mock\_RC}$}

\begin{document}
   \title{Chemical gradients in the Milky Way from the RAVE data.\\ II. Giant
stars}

   \author{
	C. Boeche\inst{1},  A. Siebert\inst{2}, 
	T. Piffl\inst{3,4}, A. Just\inst{1}, M. Steinmetz\inst{3}, 
	E. K. Grebel\inst{1}, S. Sharma\inst{5}, G. Kordopatis\inst{6}, 
	G. Gilmore\inst{6}, C. Chiappini\inst{3},
	K. Freeman\inst{7}, 
	B. K. Gibson\inst{8,9},
	U. Munari\inst{10}, 
	A. Siviero\inst{11,3}, 
	O. Bienaym\'e\inst{2},
	J.F. Navarro\inst{12}, 
	Q. A. Parker\inst{13,14,15}, W. Reid\inst{13,15},
	G. M. Seabroke\inst{16},
	F. G. Watson\inst{14}, 
	R. F. G. Wyse\inst{17},
	 T. Zwitter\inst{18}
	}          

   \offprints{corrado@ari.uni-heidelberg.de}

   \institute{
Astronomisches Rechen-Institut, Zentrum f\"ur Astronomie der 
Universit\"at Heidelberg, M\"onchhofstr. 12-14, D-69120 Heidelberg, Germany
         \and
Observatoire astronomique de Strasbourg, Universit\'e de Strasbourg, CNRS, UMR
7550, 11 rue de l'Universit\'e, F-67000 Strasbourg, France
         \and
Leibniz Institut f\"ur Astrophysik Potsdam (AIP), An der Sternwarte 16, 
D-14482 Potsdam, Germany
	\and
Rudolf Peierls Centre for Theoretical Physics, Keble Road, Oxford OX1 3NP,
UK
	\and
Sydney Institute for Astronomy, School of Physics A28, University of Sydney, NSW 2006, Australia
	\and
Institute of Astronomy, University of Cambridge, Madingley Road, Cambridge CB3 0HA, UK
        \and
 Research School of Astronomy and Astrophysics, Australian National
 University, Cotter Rd., ACT 2611 Weston, Australia 
	\and
Institute for Computational Astrophysics, Dept of Astronomy \& Physics,
Saint Mary's University, Halifax, NS, BH3 3C3, Canada
        \and
Jeremiah Horrocks Institute, University of Central Lancashire, Preston,
 PR1~2HE, United Kingdom
	\and
INAF Osservatorio Astronomico di Padova, Via dell'Osservatorio 8, Asiago I-36012, Italy
        \and
Department of Physics and Astronomy, Padova University, Vicolo
dell'Osservatorio 2, I-35122 Padova, Italy
	\and
Department of Physics and Astronomy, University of Victoria,
Victoria, BC, Canada V8P5C2
	\and
Department of Physics \& Astronomy, Macquarie University, Sydney, NSW 2109
 Australia 
	\and
Australian Astronomical Observatory, PO Box 915, North Ryde, NSW 1670, Australia
	\and
Research Centre for Astronomy, Astrophysics and
Astrophotonics, Macquarie University, Sydney, NSW 2109 Australia 
        \and
Mullard Space Science Laboratory, University College London, Holmbury St
Mary, Dorking, RH5 6NT, UK
        \and
Department of Physics and Astronomy, Johns Hopkins University, 3400 North
Charles Street, Baltimore, MD 21218, USA
        \and
Faculty of Mathematics and Physics, University of Ljubljana, Jadranska 19, SI-1000 Ljubljana, Slovenia
}


 
  \abstract
  {}
  {We provide new constraints on the chemo-dynamical models of the Milky
Way by measuring the radial and vertical chemical gradients for the elements
Mg, Al, Si, Ti, and Fe in the Galactic disc and the gradient variations
as a function of the distance from the Galactic plane (\z).}
  {We selected a sample of giant stars from the RAVE database
using the gravity criterium 1.7$<$\logg$<$2.8. 
We created a RAVE mock sample with the Galaxia code based on the Besan\c con model and
selected a corresponding mock sample to compare the model with the
observed data. We measured the radial gradients and the vertical gradients
as a function of the distance from the Galactic plane \z\ to study
their variation across the Galactic disc.}
  {The RAVE sample exhibits a negative radial gradient of
$d[Fe/H]/dR=-0.054$~dex~kpc$^{-1}$ close to the Galactic plane
($|Z|<0.4$~kpc) that becomes flatter for larger $|Z|$. Other elements follow
the same trend although with some variations from element to element. 
The mock sample has radial gradients in fair agreement with the observed
data.  The variation of the gradients with \z\ shows that the Fe radial gradient
of the RAVE sample has little change in the range $|Z|\lesssim0.6$~kpc and then flattens.
The iron vertical gradient of the RAVE sample is slightly negative close to
the Galactic plane and steepens with $|Z|$. The mock
sample exhibits an iron  vertical gradient that is always steeper than the RAVE sample.
The mock sample also shows an excess of metal-poor stars in the
[Fe/H] distributions with respect to the observed
data.  These discrepancies can be reduced by decreasing the number of thick disc
stars and increasing their average metallicity in the Besan\c con
model.}
   {}

   \titlerunning{RAVE chemical gradients in the MW}
    \authorrunning{Boeche et al.}

   \maketitle
%
\section{Introduction}\label{intro_grad}
One of the main structures
of the Milky Way is the disc, which, since the work by Gilmore \& Reid
\cite{gilmore}, has been thought to be composed of two separate components:
the thin and the thick disc. These discs are known to vary in their kinematic
and chemistry (Freeman \& Bland-Hawthorn, \citealp{freeman}), therefore they
must have experienced different star formation histories. 
Recently, the thin-thick disc model
has been questioned, since new comprehensive data do not show a clear
duality (Ivezi{\'c} et al. \citealp{ivezic}; Bovy et al. \citealp{bovy}).
The latest hypothesis suggests that the disc may be one single structure 
with kinematical and chemical features that are continuously distributed, and 
that the thin and thick discs can be seen as the two extreme tails of such a structure.
A study of the chemical gradients within the disc has the potential to shed
light on its structure and provide important 
constraints on the formation scenarios of the Galaxy.
If star formation in the
thick and thin discs occurs independently, it will be detectable through
tracing the spatial distribution of chemical abundances with reference to
the kinematic structure of the disc.\\
Several studies were previously undertaken using Cepheids, open clusters,
planetary nebulae, and turn-off stars to trace abundances (see
references in Boeche et al. \citealp{boeche13}, hereafter Paper~1). 
These studies provided radial chemical gradients ranging
from $-0.03$ to $-0.17$~dex kpc$^{-1}$, and many of them
indicate values close to $\sim-0.05$~dex kpc$^{-1}$. 
The cause of the fairly large spread among the gradients found by 
different authors may be because of the different
tracers employed. Tracers of different ages
represent the chemical evolutionary states of the Galaxy at different times (i.e.
at the time when they formed), therefore they can have significantly different gradients.
Another reason for the range of the gradient values found in 
literature may be found in local inhomogeneities because of
disrupted open clusters \cite[Montes et al.][and references
therein]{montes}, moving groups generated by resonant features due to spiral arms
(\citealp[Carlberg \& Sellwood][]{carlberg85}), the Galactic bar \cite[Minchev
\& Famaey][]{mf10}, or stellar streams due to disrupted Galactic
satellites \cite[Belokurov et al.][and reference therein]{belokurov}.\\
All of these causes of inhomogeneity can affect the chemical 
gradient measurements and make
their determination and their interpretation as result of the
star formation history of an ideally undisturbed galactic disc more challenging.
It is interesting to note how these mechanisms are also
employed to explain the kinematical heating of the thin disc or the formation of the thick disc.
For instance, Sales et al. \cite{sales} discussed the major mechanisms of thick
disc formation, including accretion from disrupted satellites (Abadi et al.
\citealp{abadi}) and heating a pre-existing disc by minor mergers (Quinn et al.
\citealp{quinn}; Villalobos \& Helmi \citealp{villalobos}). 
Kroupa \cite{kroupa} suggested that
the thick disc may be composed of stars formed in star clusters of the early thin disc
which, after their evaporation from the clusters, would acquire kinematics
that resemble the thick disc. Kinematic heating and radial mixing can also
increase the velocity dispersion of a thin disc 
(Sellwood \& Binney \citealp{sellwood02}; Ro{\v s}kar et al. 
\citealp{roskar08}; Minchev \& Famaey \citealp{mf10}, among others).
Since these mechanisms are still in action in the
present Milky Way, their effects overlap, generating a natural dispersion in the
chemical abundance distribution, and making the original causes of the chemical gradients 
difficult to disentangle.
It is worth noting that Balser et al. \cite{balser}, by
studying \ion{H}{ii} regions, found differences in
radial chemical gradients ranging from $-0.03$ to $-0.07$~dex kpc$^{-1}$
when measured at different Galactic azimuth, remarking that the abundance
distribution in the disc is not only a function of the Galactocentric
distance. Therefore, differences
in radial chemical gradients of $\sim0.04$~dex kpc$^{-1}$ between different areas 
of the disc may occur. If so, precise chemical gradient determinations
can reveal inhomogeneities among different regions of the Galaxy, but they
may not be decisive in determining the contribution of the different mechanisms 
to the Galactic disc structure.\\ 
Numerous chemical models were
developed to match the radial gradients observed.
By using different combinations of various ingredients (inside-out formation
of the disc, Chiappini et al. \citealp{chiappini}, \citealp{chiappini2001};
cosmological hydrodynamical simulations with feedback energy, Gibson et al.,
\citealp{gibson}; chemodynamical simulation with the smooth particle
hydrodynamical method, Kobayashi \& Nakasato \citealp{kobayashi}; N-body
hydrodynamical codes with chemical evolution implemented, Minchev et al.
\citealp{minchev}, \citealp{minchev14b}; chemical models with radial mixing, Sch\"onrich \& Binney,
\citealp{schoenrich}) the chemical models can cover the range of observed radial gradients
cited above (from flatter gradients like $-0.04$~dex kpc$^{-1}$ predicted by
the Chiappini et al. \citealp{chiappini} model to the steeper $-0.11$~dex
kpc$^{-1}$ of the Sch\"onrich \& Binney model).\\
Thanks to the many
spectroscopic surveys undertaken in recent years (the Geneva
Copenhagen Survey (GCS), Nordstr\"om et al.  \citealp{nordstrom04}; 
the Sloan Extension for Galactic Understanding and Exploration (SEGUE), 
Yanny et al. \citealp{yanny}; the RAdial Velocity Experiment (RAVE),
Steinmetz at al. \citealp{rave}; the Gaia-ESO survey, 
\citealp[Gilmore et al.][]{gilmore12}; the GALactic Archaeology with HERMES 
survey (GALAH), \citealp[Zucker et al.][]{zucker};
the Apache Point Observatory Galactic Evolution Experiment
survey (APOGEE), \citealp[Majewski et al.][]{majewski}; the Large sky Area Multi-Object
fiber Spectroscopic Telescope (LAMOST), \citealp[Zhao et
al.][]{zhao}; and soon Gaia, \citealp[Perryman et al.][]{perryman}), chemical abundances
of a large sample of stars were (and will be) measured for a more
comprehensive study of the Milky Way disc in the solar neighborhood and
beyond.\\

In Paper~1 we
undertook a study of the radial chemical gradients in the Milky Way
using dwarf stars that were observed and measured in the RAVE survey (Steinmetz et
al., \citealp{rave}) with the more recent stellar parameters, chemical
abundances, and distances reported in the Data Release 4 (DR4, Kordopatis et
al. \citealp{kordopatisDR4}).  
We found a radial [Fe/H] gradient
of $-0.065$~dex kpc$^{-1}$ that becomes flatter with increasing 
distance from the Galactic plane. This is in fair agreement with earlier 
and later studies (Pasquali \& Perinotto \citealp{pasquali}; Cheng et al. \citealp{cheng};
Anders et al. \citealp{anders}). The comparison with a mock sample created
with the Galaxia code (based on the Besan\c con model, 
Robin et al. \citealp{robin}) highlights an excess of
moderately low metallicity stars in the mock sample (thick disc stars) 
with respect to the observed data. Moreover, we found false positive metallicity
gradients in the (\Rg, \Zmax) plane (\Rg\ is the
guiding radius and \Zmax\ the maximum distance that the stars reach along
their Galactic orbit) caused by the lack of correlation between the metallicity 
and the kinematics of the stars in the GALAXIA mock sample.\\

In our present work, we study the radial and vertical gradients of the Galactic
disc using giant stars of the RAVE survey.  Thanks to their high
luminosity, we can probe a large volume above and below the disc, finding
similarities but also some differences with respect to the dwarf stars
studied in Paper~1.
The paper is organized as follows: in Sect.~\ref{sec_data} we describe the
data from which we extract our samples, in Sect.~\ref{error_sec} we briefly outline
the method, in Sect.~\ref{sec_analysis} we report the analysis and the
chemical gradient measured, in Sect.~\ref{discuss_sec} we discuss the
results, and we conclude in Sect.~\ref{sec_conclusion}.

\section{Data}\label{sec_data}
In the following, we summarize the characteristics of the data, which are the
same employed in Paper~1 (Boeche et al., \citealp{boeche13}). 
Because this sample was selected from the RAVE internal database in 2012,
the selection rendered fewer stars than by using the present DR4
database, which contains the complete collection of spectra observed by RAVE. 
Nonetheless, the radial velocities, stellar parameters, chemical abundances, and
distances are the same as the DR4 data release (Kordopatis et al.
\citealp{kordopatisDR4}).
The stellar atmospheric parameters, effective temperature
\temp, gravity \logg, and metallicity \met\ are measured by the new RAVE
pipeline (Kordopatis et al. \citealp{kordopatisDR4}) which makes use of the
MATrix Inversion for Spectral SynthEsis algorithm
(MATISSE) \cite[Recio-Blanco et al.][]{recio-blanco} and the
DEcision tree alGorithm for AStrophysics (DEGAS)
\cite[Bijaoui et al.][]{bijaoui}. Expected errors at S/N$\sim$50 for giant stars 
are $\sim100$~K in \temp\ and $\sim0.3$~dex for \logg. The chemical abundances
of the elements Fe, Mg, Al, Si, Ti, and Ni are derived by the RAVE chemical 
pipeline (Boeche et al. \citealp{boeche11}) with some improved features
described in Kordopatis et al. \cite{kordopatisDR4}. Errors in chemical
abundances for S/N$>$40 are estimated to be $\sim0.10-0.15$~dex for Fe, Mg, Al, and Si,
$\sim0.2$~dex for Ti, and $\sim0.25$~dex for Ni.
Proper motions are given as in the DR3 data release (Siebert et al.
\citealp{siebert}), i.e. for every star we adopted the proper motion 
with the smallest error chosen among several catalogues (Tycho2, H\o g et al.,
\citealp{hog}; the PPM-Extended catalogues PPMX and PPMXL, Roeser et al.,
\citealp{roeser2008}, \citealp{roeser2010}; the Second and Third U.S. 
Naval Observatory CCD Astrograph Catalog UCAC2 and UCAC3, Zacharias et al.,
\citealp{zacharias}). Proper motion errors vary depending on the
catalogue and the source considered. Most of the sources have average errors
of $\sim4-8$~mas~yr$^{-1}$ (Siebert et al., \citealp{siebert}). 
Distances are estimated by Binney et al. \cite{binney}. Distance errors are
estimated to lie between 20\% and 40\% in linear distance, depending on the
spectral class.\\
With these data we computed absolute positions and velocities of the stars
with respect to the Galactic centre. We integrated the Galactic orbits of
the stars with the code NEMO (Teuben, \citealp{teuben}) adopting the Galactic
potential model n.2 by Dehnen \& Binney \cite{dehnen}, which assumes
R$_\odot$=8.0 kpc, and best matches the observed properties of the Galaxy. 
Orbit integration was done for a time of 40 Gyr\footnote{A short
integration time such as one or few Gyrs would not suffice to find good approximations of the
orbital parameters. The star covers its orbit in the meridian plane after infinite time,
therefore, with a long integration time we can better approximate the
parameters $R_\mathrm{a}$, $R_\mathrm{p}$, and \Zmax at the present time.}
to extract information like apocentre
$R_\mathrm{a}$, pericentre $R_\mathrm{p}$, and maximum distance from the
Galactic plane reached by the star along its orbit \Zmax.
Using the rotation curve of the Galaxy, we converted the angular momentum into the
guiding radius \Rg.\\

From the RAVE internal database the data sample was selected
with the following constraints:
i) spectra with S/N$>$40 
ii) for stars with multiple observations we selected only the data from spectra
with highest S/N (to avoid repeated observations of the same object) 
iii) spectra with high quality flags ($Algo\_Conv=0$, 
$\chi^2<2000$, $frac>0.7$) 
iv) stars with radial velocity errors smaller than 8~km sec$^{-1}$
v) stars classified as normal stars by Matijevic
et al. \cite{matijevic}
vi) stars having [Fe/H] estimation.
With this selection we are left with 98074 RAVE stars. This sample is the
very same sample used in Paper~1 to study the chemical gradients of the dwarf stars. 
In this work, we select and study giant stars.

\subsection{The RAVE giant stars sample}\label{rave_samples_sec}
During the analysis of the data we realized that in the RAVE sample 
selected with the constraints outlined  in the previous section,
giant stars with \temp$<4250$~K were not included. 
This is due to the quality criterium $Algo\_Conv=0$,
which selects only spectra for which the DR4 pipeline code is expected to
give more accurate results thanks to the better interpolation
between the grid points of the stellar parameter space with MATISSE.
At low \temp, closer to the grid boundary and where the molecular lines
become more and more prominent, the DR4 pipeline favours the DEGAS results
over the results of MATISSE. In order to work with homogeneous data we
limited our sample to the MATISSE results.

\begin{figure}[t]
\centering 
\includegraphics[width=9cm,clip,viewport=57 491 285 679]{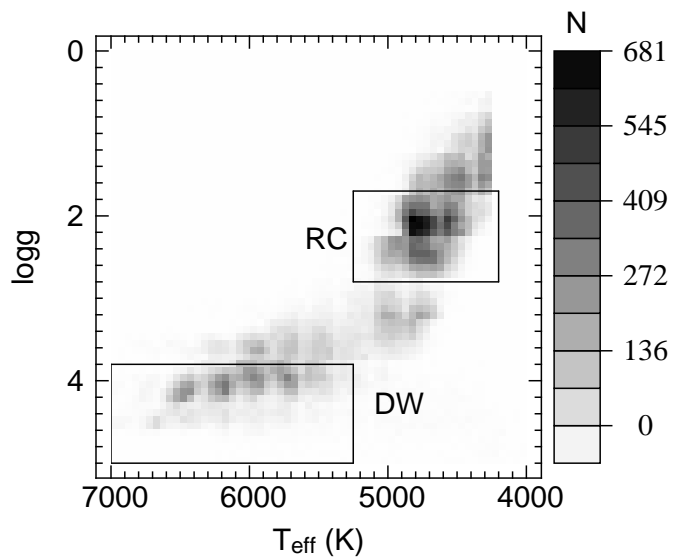}
\caption{Density distribution of our RAVE sample in the (\temp, \logg)
plane. The RAVE RC sample as selected by the
selection criteria and the dwarf stars sample (DW) as selected in paper~1
are reported.
}
\label{Teff_logg} 
\end{figure}

For stars with \logg$<1.7$ this temperature cut generates a
metallicity bias against high metallicity stars, because their isochrones
lie at lower \temp\ with respect to the low metallicity stars. 
This bias can heavily affect the chemical gradients,
therefore we decided to exclude stars with \logg$<1.7$ from our analysis.
Giant stars with \logg$>1.7$ are located at higher \temp\ and they are
not affected by this bias (see Fig.~\ref{Teff_logg}).
We selected our sample as follows:
i) effective temperature 4250$<$\temp(K)$<$5250, 
ii) gravity 1.7$<$\logg$<$2.8, and
iii) distance estimation uncertainties smaller than 30\%.
This selection leaves us with 17\,950 stars.\\
The selection criteria in \temp\ and \logg\ are represented by the rectangular set in
Fig.~\ref{Teff_logg}. Because this is the region of the red clump (RC) stars,
we call the sample ``the RC sample". In the same figure we also report the set in \temp\
and \logg\ used in Paper~1 to select the dwarf stars (DW) sample.

\begin{figure}[t]
\centering 
\includegraphics[width=9cm,clip,viewport=94 491 329 678]{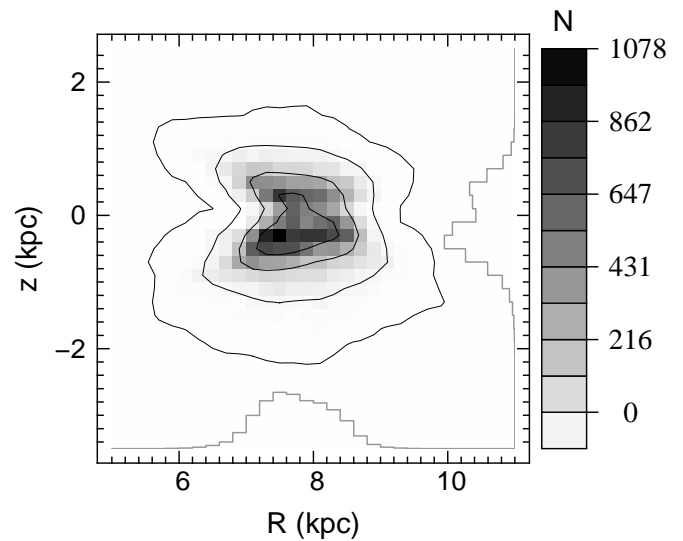}
\caption{Distribution in \R\ and \z\ for the red clump sample. 
The grey levels represent the number of stars per bin. 
The isocontours include 34\%, 68\%, and 95\% of the sample, respectively. The
histograms represent the distribution in relative numbers of the stars in
\R\ and \z.}
\label{RZgal_RC_RG} 
\end{figure}

\subsection{The mock sample} 
In order to avoid possible misinterpretations due to observational biases we
used the code GALAXIA (Sharma et al. \citealp{sharma}) to create a mock
sample to be compared with the RAVE sample.
Since GALAXIA employs the analytical
density profiles of the Besan\c con model (Robin et al.
\citealp{robin}), the mock sample is a GALAXIA realization of the Besan\c con
model but using Padova isochrones (Bertelli et al. \citealp{bertelli};
Marigo et al. \citealp{marigo}). We generated a mock sample having the same $I$ 
magnitude and colour selections as that of RAVE
stars with S/N$>$40 and flagged as normal stars by Matijevi{\v c} et al.
\cite{matijevic} (same as in Paper~1).
Out of the 97485 stars of this mock RAVE sample, 
the mock RC sample was selected with the same selection
criteria used for the RAVE RC sample outlined in Sect.~\ref{rave_samples_sec}.
The mock RC sample consists of 33\,808 stars.
This sample is numerically larger than the RAVE RC sample
because the spectrum quality selection criteria (such as $Algo\_Conv$,
$\chi^2$ and $frac$ parameters) and distance uncertainty criterium  are
applied to the RAVE sample but not to the mock sample. Like for the DW sample 
discussed in Paper~1, for the RC sample these quality selection criteria are expected to be
independent from the stellar parameters and cause no effects
on the chemical gradients.

\section{Method and error estimation}\label{error_sec}
Here we only summarize the method and error estimations, because
they are the very same applied to the DW sample and they
are fully discussed in Sect.~3 of Paper~1. 
We measured gradients by fitting the distribution of
the stars in the plane (\R,\z) and (\Rg,\Zmax) with a linear regression, and 
estimated the uncertainties with the bootstrap technique over 1000
realizations. The confidence intervals reported in the paper represent the
internal error, because undetected inhomogeneities
in the stellar samples (such as disrupted open clusters or stellar streams)
can affect the gradients in uncontrolled ways and lead to larger
uncertainties. In Paper~1, we estimated the external errors to 
be of the order of $\sim0.01$dex~kpc$^{-1}$. The Galactic orbits 
of the stars of the mock sample and the orbit parameters \Rg\ and
\Zmax\ were integrated  using the same procedure applied for the RAVE stars.
The use of kinematics-dependent parameters like \Rg\ and \Zmax\
introduces a bias that affects the estimation of the chemical gradients. As
discussed in Paper~1, stars with small \Rg\ reach the solar neighbourhood only if they are
on eccentric (kinematically hot) orbits. On average, these stars are
metal-poorer than stars having the same \Rg\ and moving on circular orbits,
which do not reach the solar neighbourhood, and are therefore missing from
the RAVE sample. This generates a bias against metal-rich stars having small
\Rg. The bias makes the gradient estimations less negative (or even
positive) than thay should be, and therefore these are estimations
to be taken with great care. In Paper~1 we had no choice but to use the
\Rg\ and \Zmax\ parameters because the spatial range in \R\ and \z\
covered by the dwarf stars is too small to permit a robust radial chemical
gradient estimate. With the RC stars we can use the \R\ and \z\
parameters because, thanks to their high luminosity, they cover a spatial volume
large enough for good chemical gradient estimations (see Fig.~\ref{RZgal_RC_RG}). 
Because of the bias that affects the gradient estimations in the (\Rg,\Zmax) plane, 
in this work we analysed the (\R,\z) plane and the results obtained using
the (\Rg,\Zmax) plane are only briefly presented for comparison with
Paper~1.\\

\subsection{Errors due to spatial distribution and distances}
The measured gradients may suffer from some systematics
due to the spatial distribution of the RAVE sample. 
In the \z\ bins the stars may be not evenly distributed 
in \R, leading to gradients that can be more affected by stars located in the
inner or outer disc. 
In the discussion of Sect.~\ref{discuss_sec} we neglected these
effects, and our comparisons refer to the mock sample in which the 
uneven spatial distribution of the stars is reproduced. For the vertical
gradients, we measured the gradients in the range
7.5$<$\R(kpc)$<$8.5 to reduce this effect as much as possible.

We also evaluated the impact of the distance errors on the gradients by
adding random and systematic errors of 30\% in distances to the mock sample.
When random errors are applied, we found negligible differences in radial 
and vertical gradients (smaller than the internal errors),
but the vertical gradient at 0.8$<|Z|\mbox{kpc}<1.2$ becomes less
steep by $\sim0.1$~dex~kpc$^{-1}$. The systematic errors induce more
consistent variations. When inflating (shrinking) the distances by 30\%  the
radial gradients show a negligible difference at \z$<0.8$~kpc, but become less
(more) positive by $\sim0.02$~dex~kpc$^{-1}$ at \z$>0.8$~kpc. The vertical
gradients flatten (steepen) by $\sim0.1$~dex~kpc$^{-1}$.

\section{Analysis and results}\label{sec_analysis}
Following the procedure adopted in Paper~1, we measured the chemical
gradients of the RC sample, dividing it into different sub-samples with different
distances $|Z|$ from the Galactic plane and then investigate how the gradients
change with \z. Unlike the dwarf stars studied in paper~1, 
the RC sample cover a volume large enough to study the gradients
in the (\R,\z) plane at four (instead of three) \z\ ranges.
The RAVE RC sample stars do not lie beyond $|Z|$=2.5~kpc, therefore the 
choosen $|Z|$ bins extend up to 2~kpc, which excludes only 2 stars.\\
We also investigate the difference in chemical gradients of
the $\alpha$-enhanced stars from the non $\alpha$-enhanced stars. The
selection criteria of this further subdivision are outlined in
Sect.~\ref{sec_alpha_poor_rich}.

\subsection{Radial chemical gradients for the RAVE RC sample}\label{RC_grad_sec}
We divided the RAVE RC sample in
four sub-samples: $0.0<|Z|$ (kpc)$\leq 0.4$ (named \RCa\ sample),
$0.4<|Z|$ (kpc)$\leq 0.8$ (the \RCb\ sample), 
$0.8<|Z|$ (kpc)$\leq 1.2$ (the \RCc\ sample), 
and $1.2<|Z|$ (kpc)$<2.0$ (the \RCd\ sample).
We omitted a few outliers located at \R$>$9.5~kpc and \R$<$4.5~kpc.
With these samples we found that the radial gradients become progressively less
negative with \z\ (see Fig.~\ref{fitting_RCG_RZgal}). For iron the
gradients are:\\

\noindent
$\frac{d [Fe/H]}{d R}
\mbox{\RCa}=-0.054\pm0.004$ dex kpc$^{-1}$
(8\,459 stars);\\
$\frac{d [Fe/H]}{d R}
\mbox{\RCb}=-0.039\pm0.004$ dex kpc$^{-1}$
(7\,651 stars);\\
$\frac{d [Fe/H]}{d R}
\mbox{\RCc}=-0.011\pm0.008$ dex kpc$^{-1}$
(1\,532 stars); and\\
$\frac{d [Fe/H]}{d R}
\mbox{\RCd}=+0.047\pm0.018$ dex kpc$^{-1}$
(283 stars).\\
\noindent

We also measured the chemical gradients for the elements Mg, Al, Si, and Ti
and their abundance relative to Fe.
The results are reported in Tabs.~\ref{tab_RC_XH_grad_R} and
\ref{tab_RC_XFe_grad_R}.\\
To test the robustness of the results and to compare them with the results in
Paper~1 we also measured the chemical gradients in the (\Rg,\Zmax) plane. To
do so we selected four sub-samples as a function of \Zmax\ with boundaries of
$0.0<Z_\mathrm{max}$ (kpc)$\leq 0.4$, $0.4<Z_\mathrm{max}$ (kpc)$\leq 0.8$,
$0.8<Z_\mathrm{max}$ (kpc)$\leq 1.2$, and $1.2<Z_\mathrm{max}$ (kpc)$\leq
10.0$ and we found iron gradients of $d[Fe/H]/dR_g=-0.029\pm0.003$, $-0.014\pm0.002$,
$+0.000\pm0.003$, and $0.029\pm0.003$ dex~kpc$^{-1}$, respectively.
The flatter gradients are due to the bias discussed in Sect.~\ref{error_sec}.

\subsection{Radial chemical gradients for the mock RC samples}
By applying the same cuts in \z\ seen before, for the mock RC sample the Fe
radial gradients are (see Fig.~\ref{fitting_RCG_Mock_RZgal}):\\

\noindent
$\frac{d [Fe/H]}{d R}
\mbox{\RCma}=-0.049\pm0.005$ dex kpc$^{-1}$
(15\,524 stars);\\
$\frac{d [Fe/H]}{d R}
\mbox{\RCmb}=-0.019\pm0.005$ dex kpc$^{-1}$
(13\,304 stars);\\
$\frac{d [Fe/H]}{d R}
\mbox{\RCmc}=+0.030\pm0.009$ dex kpc$^{-1}$
(3\,599 stars); and\\
$\frac{d [Fe/H]}{d R}
\mbox{\RCmd}=+0.061\pm0.012$ dex kpc$^{-1}$
(1\,189 stars).\\
\noindent

For the mock sample there are no other elements for which we can measure the
gradients because the model provides only iron abundance.
We measured the gradients for the mock RC sample in the (\Rg,
\Zmax) plane and we found that they are always positive, similar to what we 
found in Paper~1 for dwarf stars. For the intervals
$0.0<Z_\mathrm{max}$ (kpc)$\leq 0.4$,
$0.4<Z_\mathrm{max}$ (kpc)$\leq 0.8$,
$0.8<Z_\mathrm{max}$ (kpc)$\leq 1.2$, and $1.2<Z_\mathrm{max}$ (kpc)$\leq
10.0$ the iron gradients for the RC mock sample are $d[Fe/H]/dR_g=+0.051\pm0.003$,
$+0.057\pm0.002$, $+0.068\pm0.004$, and $+0.064\pm0.007$ dex~kpc$^{-1}$.
The comparison with the RAVE RC sample is discussed in
Sect.~\ref{sec_comp_mock}.

\subsection{Vertical chemical gradients for the RAVE RC sample}
The RAVE vertical gradients are not constant but exhibit
variations as a function of \z. We investigate these variations by following
the continuous abundances variation along \z\ (see
Sect.~\ref{sec_smooth}) and by measuring them in four \z\ intervals (the
same intervals as employed for the radial gradients).
Because the RAVE stars cover a volume that is roughly (non-symmetric) cone-shaped, 
their \R\ distribution depends on their \z. This can affect the estimation of
the vertical gradient because the average abundances at a given \z\ may refer to
different \R. To limit this bias as much as possible, we measured the vertical 
gradients at \R$\sim8$~kpc by considering only stars within the interval
7.5$\le$\R(kpc)$<$8.5. 
We summarize here the vertical gradients for the iron abundance of the RAVE
RC sample:\\

\noindent
$\frac{d [Fe/H]}{d Z}
\mbox{\RCa}=-0.050\pm0.027$ dex kpc$^{-1}$
(5\,903 stars);\\
$\frac{d [Fe/H]}{d Z}
\mbox{\RCb}=-0.087\pm0.030$ dex kpc$^{-1}$
(3\,815 stars);\\
$\frac{d [Fe/H]}{d Z}
\mbox{\RCc}=-0.148\pm0.073$ dex kpc$^{-1}$
(661 stars); and\\
$\frac{d [Fe/H]}{d Z}
\mbox{\RCd}=-0.199\pm0.070$ dex kpc$^{-1}$
(129 stars).\\
\noindent

When the iron vertical gradient is measured over the whole
sample in the $0.0<|Z|<2.0$~kpc range, it amounts to
$\frac{d [Fe/H]}{d Z}=-0.112\pm0.007$ dex kpc$^{-1}$
(10\,511 stars).
The vertical gradients for the individual elements
are reported in Tab.~\ref{tab_RC_XH_grad_z}.

\subsection{Vertical chemical gradients for the mock RC sample}
For the mock sample we repeated the same procedure used to measure the
vertical gradient of the RAVE RC sample. They are significantly steeper than in the 
observational data, and they are:\\

\noindent
$\frac{d [Fe/H]}{d Z}
\mbox{\RCma}=-0.373\pm0.031$ dex kpc$^{-1}$
(10\,776 stars);\\
$\frac{d [Fe/H]}{d Z}
\mbox{\RCmb}=-0.409\pm0.038$ dex kpc$^{-1}$
(6\,224 stars);\\
$\frac{d [Fe/H]}{d Z}
\mbox{\RCmc}=-0.536\pm0.085$ dex kpc$^{-1}$
(1\,440 stars); and\\
$\frac{d [Fe/H]}{d Z}
\mbox{\RCmd}=-0.338\pm0.083$ dex kpc$^{-1}$
(471 stars).\\
\noindent

When the vertical gradient is measured over the whole
sample in the $0.0<|Z|<2.0$~kpc range, this is
$\frac{d [Fe/H]}{d Z}=-0.400\pm0.008$ dex kpc$^{-1}$
(18\,919 stars).
For comparison this is also reported in Tab.~\ref{tab_RC_XH_grad_z}.

\begin{figure}[t]
\centering 
\includegraphics[width=9cm,clip,viewport=25 25 222 132]{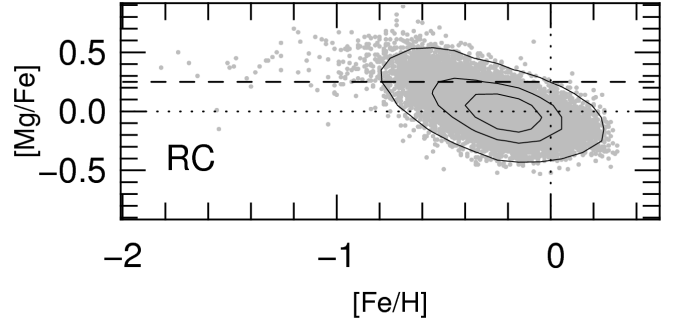}
\caption{Distribution of the RAVE RC stars in the ([Mg/Fe],[Fe/H]) plane. 
The isocountours holds 34\%, 68\%, and 95\% of the stars. The dashed line
at [Mg/Fe]=+0.2~dex separate the $\alpha$-rich stars (above the line) from the
$\alpha$-poor stars (below the line).
}
\label{MgFe_FeH} 
\end{figure}

\subsection{Chemical gradients of the $\alpha$-poor and
$\alpha$-rich stars}\label{sec_alpha_poor_rich}

Thin- and thick-disc stars (as usually defined in the frame of the 
thin- thick-disc dichotomy) chemically differ
from their enhancement in $\alpha$ elements with respect to iron. Recently,
some authors (Schlesinger et al., \citealp{schlesinger}; Hayden et al. \citealp{hayden};
Lee et al. \citealp{lee}; Cheng et al. \citealp{cheng_b}) analyzed their samples by
dividing them in $\alpha$-enhanced and non $\alpha$-enhanced stars in the
attempt to highlight their different behaviour. For comparison purposes
we follow their example and divide the RAVE RC sample as a function of the magnesium enhancement
with respect to iron [Mg/Fe]. The use of other $\alpha$ elements like Si or Ti
is not appropriate because their behaviour
looks peculiar (this is reported in Sect.~\ref{sec_discuss_RAVE}). 
Because we do not see bimodality in the chemistry distribution of the
RAVE sample (see Fig.~\ref{MgFe_FeH}) we made 
few attempts to separate these two sub-samples by cutting at [Mg/Fe]=0.15, 0.20, and 0.25~dex.
We noticed that for progressively higher [Mg/Fe] the $\alpha$-enhanced sample
can be distinguished from the non $\alpha$-enhanced sample for the different gradients,
although the smaller size of the $\alpha$-enhenced sample obtained with higher [Mg/Fe]
cuts delivers a poorer statistic. To emphasize this difference in gradients, we decided to cut at
[Mg/Fe]=0.25~dex and we call $\alpha$-rich those stars with [Mg/Fe]$>$0.25~dex and 
$\alpha$-poor those with [Mg/Fe]$\le$0.25~dex. 
With this further sub-division we perform the radial
and vertical gradients measurements with the same procedure used in the previous sections.
Here we briefly summarize the results for iron.\\

The iron radial gradients for the $\alpha$-poor stars are\\

\noindent
$\frac{d [Fe/H]}{d R}
\mbox{\RCa}=-0.054\pm0.004$ dex kpc$^{-1}$
(7\,971 stars);\\
$\frac{d [Fe/H]}{d R}
\mbox{\RCb}=-0.045\pm0.004$ dex kpc$^{-1}$
(6\,874 stars);\\
$\frac{d [Fe/H]}{d R}
\mbox{\RCc}=-0.026\pm0.008$ dex kpc$^{-1}$
(1\,256 stars); and\\
$\frac{d [Fe/H]}{d R}
\mbox{\RCd}=+0.040\pm0.019$ dex kpc$^{-1}$
(197 stars),\\
\noindent

while for the $\alpha$-rich stars they are\\

\noindent
$\frac{d [Fe/H]}{d R}
\mbox{\RCa}=-0.010\pm0.018$ dex kpc$^{-1}$
(488 stars);\\
$\frac{d [Fe/H]}{d R}
\mbox{\RCb}=+0.009\pm0.011$ dex kpc$^{-1}$
(777);\\
$\frac{d [Fe/H]}{d R}
\mbox{\RCc}=-0.002\pm0.014$ dex kpc$^{-1}$
(276 stars); and\\
$\frac{d [Fe/H]}{d R}
\mbox{\RCd}=+0.047\pm0.028$ dex kpc$^{-1}$
(86 stars).\\
\noindent

The iron vertical gradients for the $\alpha$-poor stars are\\

\noindent
$\frac{d [Fe/H]}{d Z}
\mbox{\RCa}=+0.004\pm0.025$ dex kpc$^{-1}$
(5\,566 stars);\\
$\frac{d [Fe/H]}{d Z}
\mbox{\RCb}=-0.028\pm0.028$ dex kpc$^{-1}$
(3\,461 stars);\\
$\frac{d [Fe/H]}{d Z}
\mbox{\RCc}=-0.085\pm0.076$ dex kpc$^{-1}$
(546 stars); and\\
$\frac{d [Fe/H]}{d Z}
\mbox{\RCd}=-0.141\pm0.090$ dex kpc$^{-1}$
(87 stars).\\
\noindent

The vertical gradient measured over the whole $\alpha$-poor
sample in the $0.0<|Z|<2.0$~kpc range is
$\frac{d [Fe/H]}{d Z}=-0.067\pm0.008$ dex kpc$^{-1}$
(9\,663 stars).\\

For the $\alpha$-rich stars, the vertical gradients are\\

\noindent
$\frac{d [Fe/H]}{d Z}
\mbox{\RCa}=-0.312\pm0.113$ dex kpc$^{-1}$
(337 stars);\\
$\frac{d [Fe/H]}{d Z}
\mbox{\RCb}=+0.015\pm0.092$ dex kpc$^{-1}$
(354 stars);\\
$\frac{d [Fe/H]}{d Z}
\mbox{\RCc}=+0.141\pm0.120$ dex kpc$^{-1}$
(115 stars); and\\
$\frac{d [Fe/H]}{d Z}
\mbox{\RCd}=-0.073\pm0.088$ dex kpc$^{-1}$
(42 stars).\\

\noindent
The vertical gradient measured over the whole $\alpha$-rich
sample in the $0.0<|Z|<2.0$~kpc range is
$\frac{d [Fe/H]}{d Z}=-0.021\pm0.017$ dex kpc$^{-1}$
(848 stars).
The results for the other elements are reported in 
Tabs.~\ref{tab_a_samp_grad_R} and \ref{tab_a_samp_grad_z}.

\subsection{Gradient estimates with moving box car}\label{sec_smooth}
Here we want to investigate how the radial chemical gradients change with \z.
We select sub-samples of stars that
lie in an interval 0.2~kpc wide and, starting from \z=$-2.0$~kpc, 
we shift this interval of 0.2~kpc each step up to \z=$+2.0$~kpc and measure the
gradient $d[X/H]/dR$ at every step. We impose the condition that
the interval must contain no less than
1000 stars.  If the interval contains less than 1000 stars the interval width
increases until this number is reached, with a width maximum limit of 0.6~kpc.
We avoid the interval between \z=$\pm0.1$~kpc because of the small range in \R\ 
covered by the RAVE sample in this region, and because of its scarcity of points.
This procedure was applied to the five RAVE element abundances of the RAVE RC
sample, and to the iron abundance of the mock RC sample. 
For the RAVE sample this procedure was also applied to the 
abundances relative to iron [X/Fe].
We also want to see how the
median abundances and abundances relative to iron change with \z\ 
(i.e., the vertical gradients). 
We repeated the same procedure by
measuring the median abundance of the stars contained in a \z\ interval 0.2~kpc wide
and moving the interval in 0.2~kpc steps from \z=$-2.0$~kpc to
\z=$+2.0$~kpc. Here we imposed a lower limit of 20 stars per
interval. If this limit is not reached, the interval width is increased 
up to a maximum of 0.6~kpc. 
The results for the RAVE and the mock RC samples are illustrated 
in Fig.~\ref{grad_boxcar_RCG} to Fig.~\ref{grad_boxcar_RCG_XFe_R8}.

\begin{figure*}[t]
\begin{minipage}[t]{9cm}
\includegraphics[width=9cm,clip,viewport= 21 27 405 323]{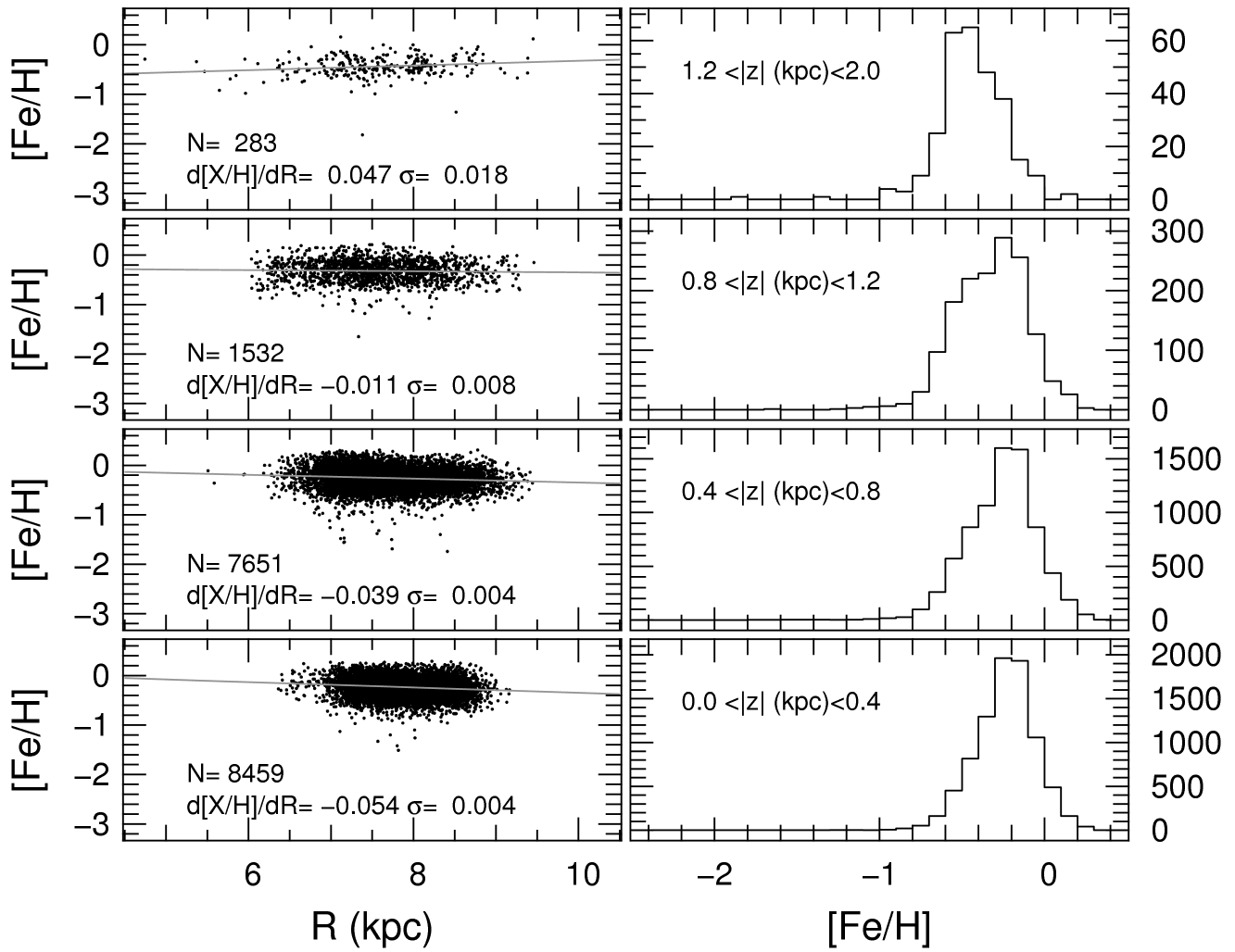}
\caption{Distribution of the RAVE RC sample 
at different \z\ intervals in the (\R,\z) plane (left panels)  and
metallicity distributions (right panels).}
\label{fitting_RCG_RZgal} 
\end{minipage}
\hfill
\begin{minipage}[t]{9cm}
\includegraphics[width=9cm,clip,viewport= 21 27 405 323]{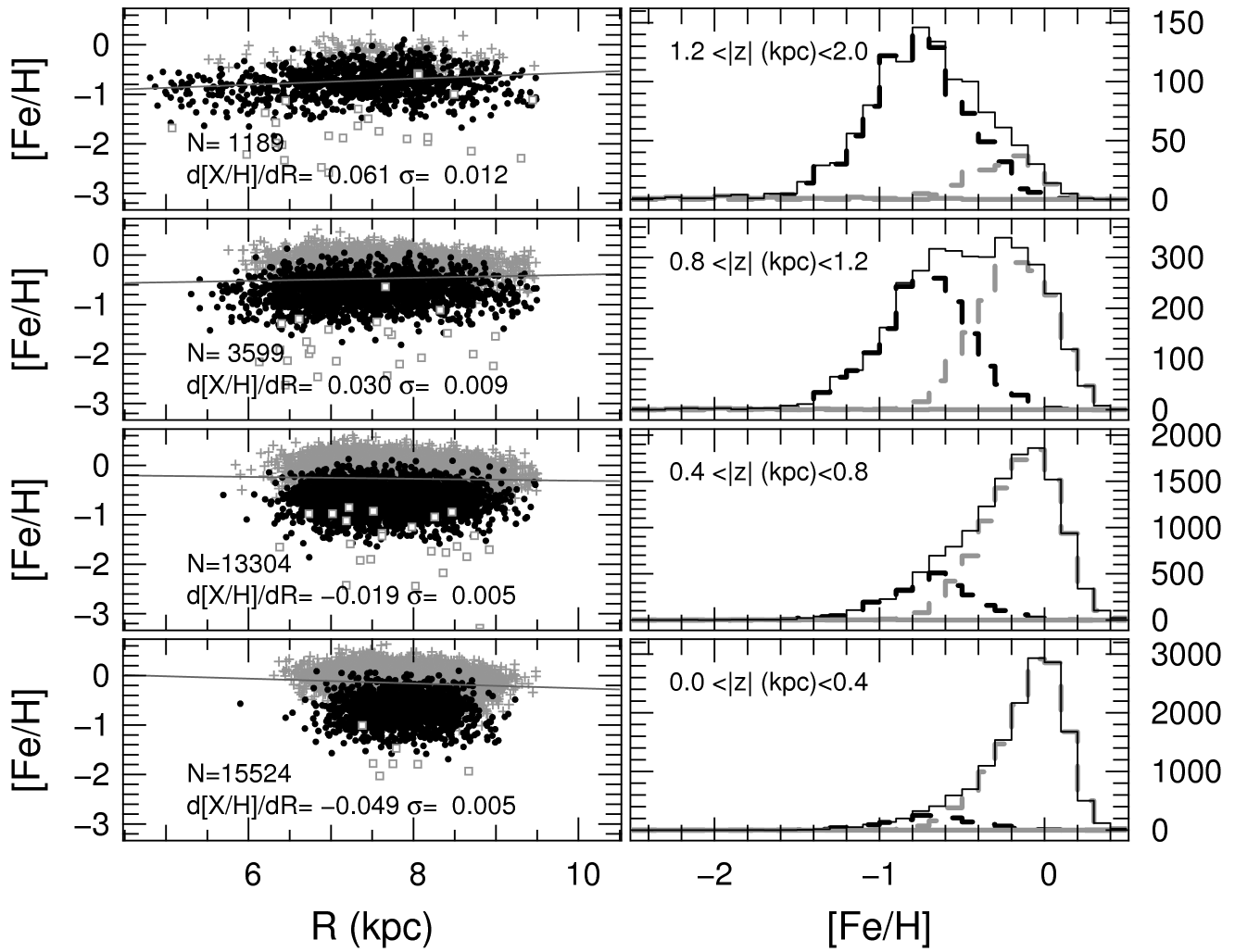}
\caption{As in Fig.~\ref{fitting_RCG_RZgal},
but for the mock sample. Grey plus symbols, black points and grey squares
represent thin, thick, and halo stars.}
\label{fitting_RCG_Mock_RZgal} 
\end{minipage}
\end{figure*}

\section{Discussion}\label{discuss_sec}
The comparison between the RAVE RC sample and the mock RC sample 
in the (\Rg, \Zmax) plane brings us to the same consideration reported 
in Paper~1 for the dwarfs stars sample:
when kinematical parameters like \Rg\ and \Zmax\ are used to estimate the
radial gradients, the mock sample
yields unrealistic positive gradients because the Fe abundances
of the stars in the model are assigned as a function of their spatial
position, disregarding their kinematics. As a consequence, the thin disk stars (which
have a radial gradient of $d[Fe/H]/dR=-0.07$dex~kpc$^{-1}$ 
in the Besan\c con model) have zero gradient in the (\Rg,\Zmax) plane 
because their position in \R\ is re-distributed in \Rg\ independent of 
their metallicity. On the other hand, the thick disc stars (which are metal
poorer and have more eccentric orbits with respect to the thin disc stars)
lie to the lower-left corner of the (\Rg,\Zmax) plane. Thin and thick disc
stars together generate an apparent positive gradient that is not real.
Such a bias is also visible in real data with a flattening of the gradients,
as we saw for the dwarf stars in Sect.~6.1 of Paper~1.
Therefore, in the following we mainly discuss the gradients in the (\R,\z)
plane.\\
At the beginning of our analysis, we divided the $|Z|$ ranges into four bins.
Because the Galactic disc is expected to be roughly symmetric with respect
to \z=0, the
choice of bins in $|Z|$ increases the number of stars per bin and makes the
statistic more robust. 
Nonetheless, we decided to study at higher resolution how
the gradient changes with \z\ by using smaller bins of 0.2~kpc width. The two analyses
are consistent with each other and the higher resolution allows us to determine if
(and where) possible transitions occur in the vertical structure of the disc.

\subsection{Chemical gradients of the RAVE RC
sample}\label{sec_discuss_RAVE}
Tabs.~\ref{tab_RC_XH_grad_R} to \ref{tab_a_samp_grad_z} hold a considerable
amount of information that is not always straightforward to interpret.
In the following we avoid discussing every chemical gradient measured, and 
we focus only on the main outcomes of our measurements. A 
detailed comparison with models (necessary to fully understand all the
reported gradients) is left to a future work.
Here we summarize the main results for the RAVE RC sample:
\begin{itemize}
\item the radial gradients are negative and
become progressively flatter with growing $|Z|$;
\item vertical gradients are negative
and become progressively steeper with growing $|Z|$;
\item the negative radial gradients observed are driven by the $\alpha$-poor
stars, while the $\alpha$-rich stars have radial gradients consistent with zero;
\item the vertical gradients of the $\alpha$-poor stars are consistent with zero
for $|Z|<0.8$~kpc and are slightly negative beyond 0.8~kpc;
\item the vertical gradients of the $\alpha$-rich stars are negative for
$|Z|<0.4$~kpc and consistent with zero for $|Z|>0.4$~kpc;
\item the radial gradients of the abundances relative to iron have small
positive or zero values up to $|Z|\sim0.8$~kpc and then become negative.
\end{itemize}

These results have general validity, although
a closer look at the gradient values reveals that there are differences
between the elements, and that some elements behave
alike. In fact, Si behaves like Fe, Al
behaves like Mg, and Ti behaves differently from all the others.
The similar behaviour of Si and Fe is unexpected, since Si is an $\alpha$
element and it should behave like Mg. The peculiar behaviour of Ti is
equally singular.
A future detailed comparison with chemo-dynamical models
will help to address the reasons behind such differences and similarities.\\

In the proximity of the Galactic plane ($|Z|<0.4$~kpc) 
the $\alpha$-poor and $\alpha$-rich samples have a remarkable behaviour.
The chemistry and
spatial location of the $\alpha$-poor stars would identify them as thin disc
stars and we would expect
that they would drive the vertical gradient at $|Z|<0.4$~kpc. Unlike the
expectations, their iron vertical gradient is consistent with zero
($+0.004\pm0.025$~dex~kpc$^{-1}$) and this is in striking contrast with
the mock sample ($-0.373\pm0.031$~dex~kpc$^{-1}$). This suggests a moderate 
chemical vertical homogeneity in the proximity of the Galactic disc and 
it may be a sign of weak correlation between metallicity and vertical 
velocity dispersion.\\ 
On the other hand, the $\alpha$-rich stars at $|Z|<0.4$ (which we would identify as
thick disc stars close to the Galactic plane) exhibit a negative iron
vertical gradient ($-0.312\pm0.113$~dex~kpc$^{-1}$), revealing a correlation
between metallicity and distance from the Galactic plane at small \z.
This reflects the result of Minchev et al. \cite{minchev14} who, using
RAVE data, found that
the vertical velocity dispersion of the giant stars increases with [Mg/Fe]
and suddenly decreases for [Mg/Fe]$\gtrsim$0.4~dex. These high Mg-enhanced
stars clump close to the Galactic plane and, according to Minchev et al. 
\cite{minchev14}, they would have migrated from the inner disc as result 
of massive mergers in the early history of the Galaxy.\\

\subsection{Comparison between the RAVE and the mock RC samples}\label{sec_comp_mock}
\begin{figure*}[t]
\begin{minipage}[t]{9cm}
\includegraphics[width=9cm,clip,viewport=93 289 296 664]{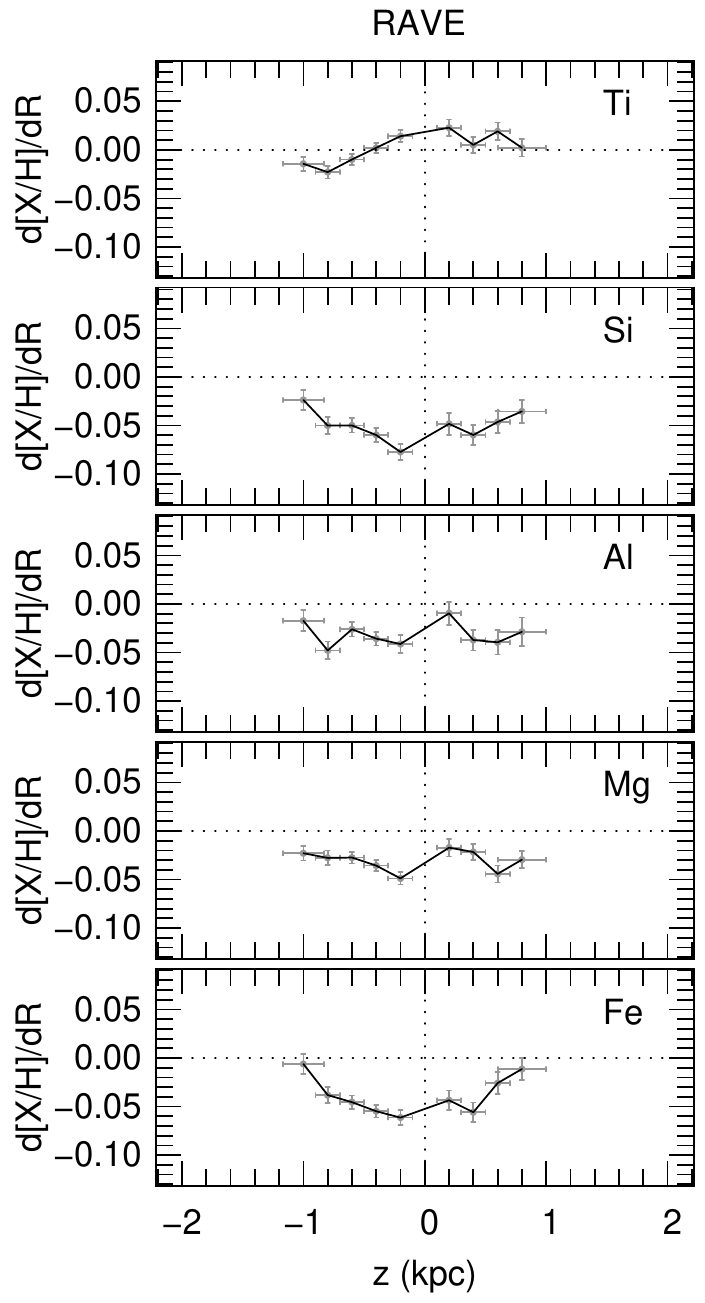}
\includegraphics[width=9cm,clip,viewport=94 290 295 402]{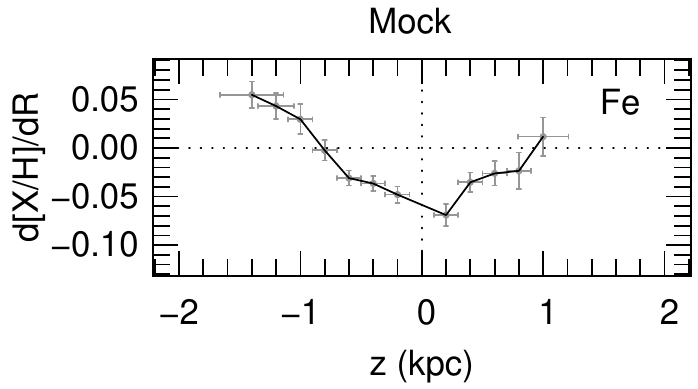}
\caption{Radial gradients for the RAVE RC sample (top panels) and the RC mock
sample (bottom panel) as a function of the distance from the Galactic plane
\z. The horizontal bars represent the \z\ interval taken to measure the
gradient $d[X/H]/dR$, while the vertical bars represent the uncertainty
(68\% confidence) of the gradient estimates.}
\label{grad_boxcar_RCG} 
\end{minipage}
\hfill
\begin{minipage}[t]{9cm}
\includegraphics[width=9cm,clip,viewport=25 28 222 401]{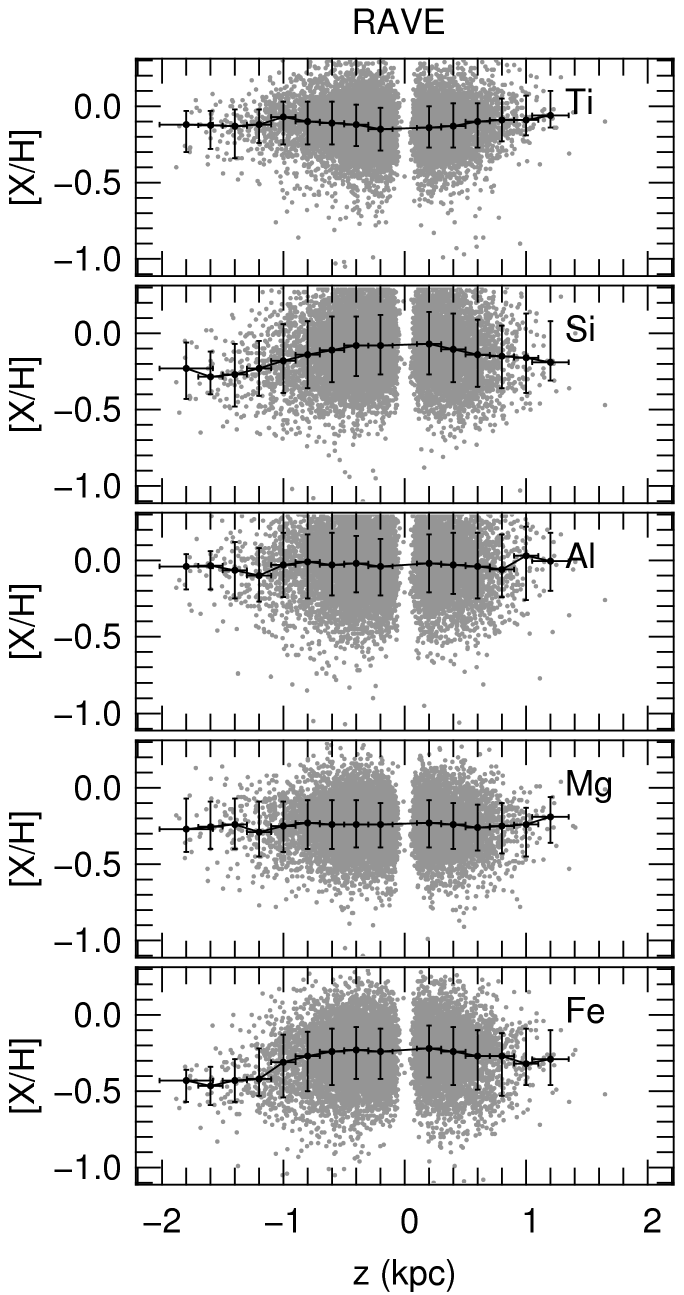}
\includegraphics[width=9cm,clip,viewport=24 27 222 140]{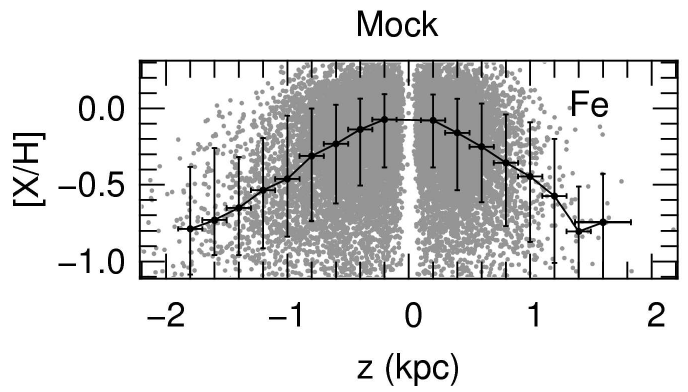}
\caption{Abundances of the RAVE RC sample (top panels) and RC mock sample
(bottom panel) as a function of \z\ in the range 7.5$<$\R(kpc)$<$8.5. The black points represent
the median abundance of the samples in step of 0.2~kpc, while the vertical
bars contain 68\% of the stars in the bin. 
}
\label{grad_boxcar_RCG_XH_R8}
\end{minipage}
\end{figure*}

\begin{figure*}[t]
\begin{minipage}[t]{9cm}
\includegraphics[width=9cm,clip,viewport=95 289 295 596]{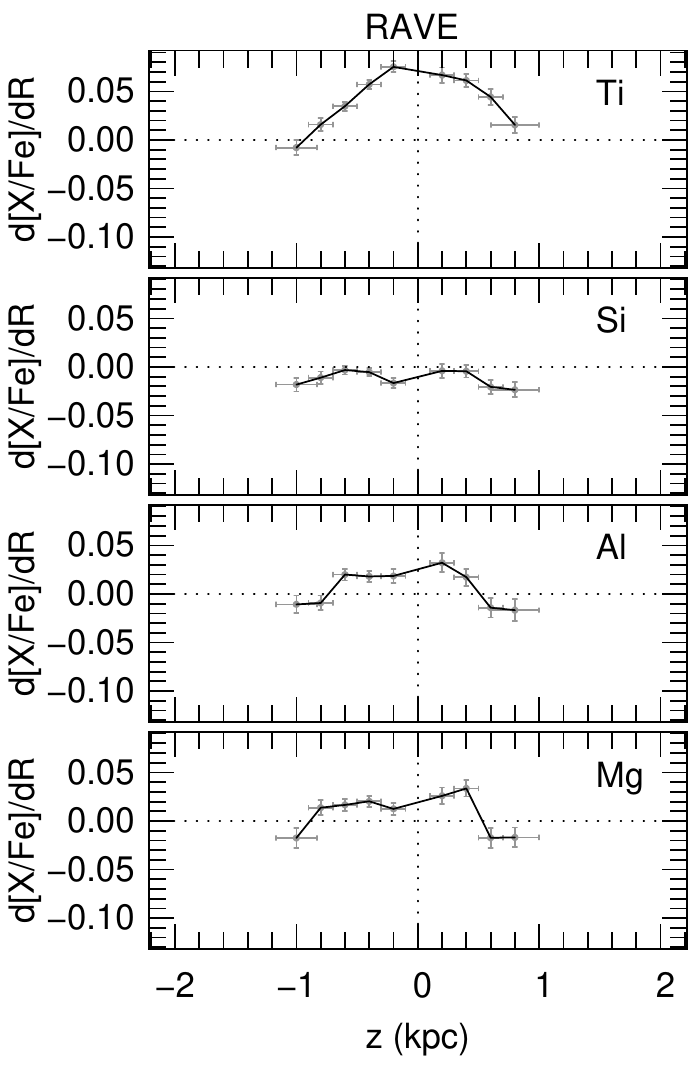}
\caption{As in Fig.~\ref{grad_boxcar_RCG} but for the elemental abundance
ratio [X/Fe].}
\label{grad_boxcar_RCG_dRXFe} 
\end{minipage}
\hfill
\begin{minipage}[t]{9cm}
\includegraphics[width=9cm,clip,viewport=25 28 221 336]{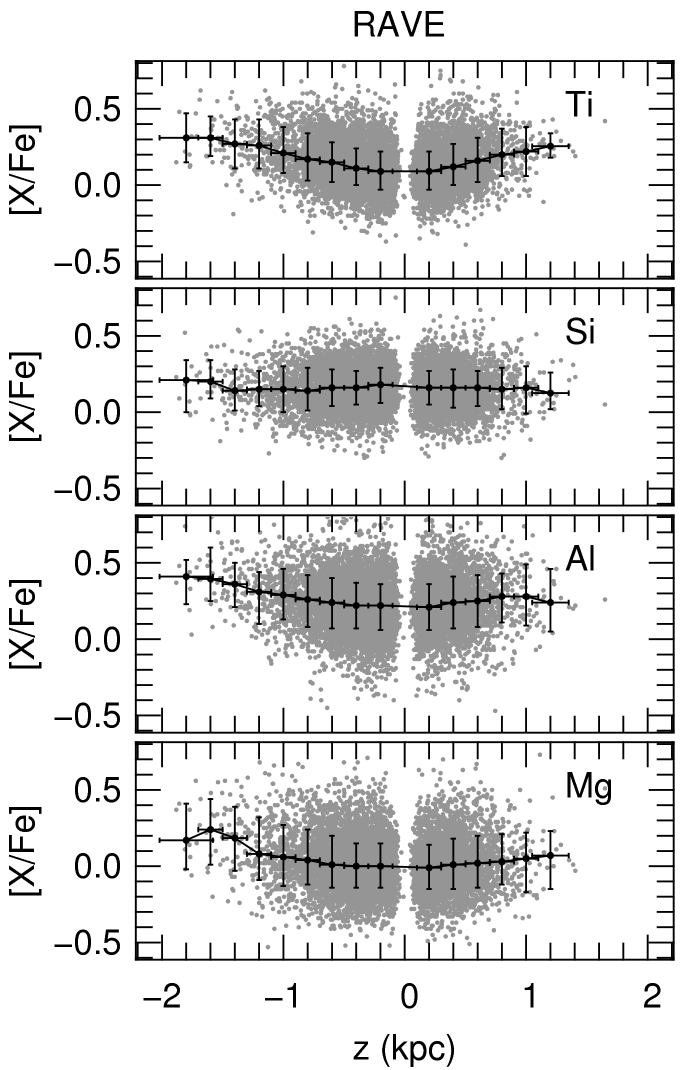}
\caption{As in Fig.~\ref{grad_boxcar_RCG_XH_R8} but for the elemental abundance
ratio [X/Fe].}
\label{grad_boxcar_RCG_XFe_R8} 
\end{minipage}
\end{figure*}

At $|Z|<$0.4~kpc the mock RC sample exhibits an iron radial gradient 
in good agreement with the RAVE RC sample and, for both samples, the iron radial
gradient becomes flatter with $|Z|$ (see Tab.~\ref{tab_RC_XH_grad_R}, 
Figs.~\ref{fitting_RCG_RZgal} and ~\ref{fitting_RCG_Mock_RZgal}).
For the stars at larger $|Z|$, the radial iron gradient becomes positive for both the RAVE
and the mock samples. This positive gradient may be due to the spatial
distribution of the RAVE stars and not to an intrinsic positive radial gradient
of the disc at that $|Z|$ range.
Significant differences between the RAVE RC sample and the mock RC
sample can be found in the vertical gradients. The iron vertical gradients
in the mock RC sample are much steeper than in the RAVE sample (see
Tab.~\ref{tab_RC_XH_grad_z} and Fig.~\ref{grad_boxcar_RCG_XH_R8}). We
explain this by comparing the right panels of Figs.~\ref{fitting_RCG_RZgal} and
\ref{fitting_RCG_Mock_RZgal}. While for the RAVE RC
sample the mode of the [Fe/H] distribution moves from [Fe/H]$\sim$-0.2~dex
to [Fe/H]$\sim$-0.5~dex in the range $0.0<|Z|$(kpc)$<2.0$, for the mock RC sample 
the mode of the [Fe/H] distribution spans $\sim0.8$~dex in the same $|Z|$
range. The grey and black dashed histograms in the right panels of
Fig.~\ref{fitting_RCG_Mock_RZgal} (representing the thick- and thin-disc
stars, respectively) clearly show that the shift of the [Fe/H] mode is due
to the thick-disc that takes over the thin-disc with growing $|Z|$.
This means that the steep vertical metallicity
observed in the mock RC sample is caused by the overlap of
the thin- and the thick-disc, which have different average metallicities
and scale heights. 
The difference in average metallicity of the two discs 
also explains the broader [Fe/H] distribution of the RC mock sample with 
respect to the RAVE RC sample.
This interpretation is also supported by the analysis of the $\alpha$-poor and $\alpha$-rich
samples. The iron vertical gradients of the $\alpha$-poor and the $\alpha$-rich samples
are consistent with zero or slightly negative values\footnote{The negative values
have low significance, because they 
are consistent with zero inside 1-2$\sigma$ (Tab.~\ref{tab_a_samp_grad_z}). Here we also neglect the 
negative value shown by the $\alpha$-rich sample at
$|Z|<0.4$, which we discussed in the previous section.}. 
This suggests that the vertical gradients seen in the RAVE RC sample can also be explained
by the superposition of two populations with zero or shallow vertical gradients
having different metallicities and scale heights: one with small scale-height 
and metal-rich and the other with large scale-height and metal poor.\\
The mock RC sample would find a better agreement with the RAVE data
if the average metallicity of the model's thick-disc was
shifted from $-0.78$~dex to $\sim-0.5$~dex.

\begin{table*}[t]
\caption[]{Radial abundance gradients measured in the RC RAVE sample for Fe, Mg, Al,
 Si, and Ti  expressed as dex kpc$^{-1}$ for four
ranges of $|Z|$. For comparison, in the last column, the vertical [Fe/H]
gradients of the mock sample are reported. Uncertainties of 68\% confidence are
obtained with the bootstrap method and represent the internal errors. 
}
\label{tab_RC_XH_grad_R}
\vskip 0.3cm
\centering
\begin{tabular}{l|ccccc|c}
\hline
\noalign{\smallskip}
        &  $\frac{d[Fe/H]}{dR}$  & $\frac{d[Mg/H]}{dR}$  & $\frac{d[Al/H]}{dR}$
& $\frac{d[Si/H]}{dR}$ & $\frac{d[Ti/H]}{dR}$ & $\frac{d[Fe/H]}{dR}$ (mock)\\
\noalign{\smallskip}
\hline
\noalign{\smallskip}
$0.0<|Z|$ (kpc)$\leq 0.4$ &$-0.054\pm0.004$&$-0.034\pm0.004$&$-0.035\pm0.005$ &$-0.064\pm0.005$ & $+0.008\pm0.004$& $-0.049\pm0.006$ \\
$0.4<|Z|$ (kpc)$\leq 0.8$ &$-0.039\pm0.004$&$-0.031\pm0.004$&$-0.032\pm0.005$ &$-0.046\pm0.004$ & $-0.005\pm0.003$& $-0.019\pm0.005$\\
$0.8<|Z|$ (kpc)$\leq 1.2$ &$-0.011\pm0.008$&$-0.023\pm0.007$&$-0.027\pm0.009$ & $-0.028\pm0.008$ & $-0.015\pm0.006$& $+0.030\pm0.009$\\
$1.2<|Z|$ (kpc)$<2.0$     &$+0.047\pm0.018$&$+0.025\pm0.015$&$+0.060\pm0.022$ & $+0.009\pm0.018$ & $+0.032\pm0.017$& $+0.061\pm0.012$\\
\noalign{\smallskip}
\hline
\end{tabular}

\end{table*}

\begin{table*}[t]
\caption[]{As Table~\ref{tab_RC_XH_grad_R}, but for relative abundances [X/Fe].}
\label{tab_RC_XFe_grad_R}
\vskip 0.3cm
\centering
\begin{tabular}{l|cccccc}
\hline
\noalign{\smallskip}
        &  & $\frac{d[Mg/Fe]}{dR}$ & 
$\frac{d[Al/Fe]}{dR}$ & $\frac{d[Si/Fe]}{dR}$ & $\frac{d[Ti/Fe]}{dR}$ \\
\noalign{\smallskip}
\hline
\noalign{\smallskip}
$0.0<|Z|$ (kpc)$\leq 0.4$ &       &$+0.020\pm0.004$&$+0.019\pm0.004$&$-0.009\pm0.003$&$+0.063\pm0.003$\\
$0.4<|Z|$ (kpc)$\leq 0.8$ &       &$+0.009\pm0.004$&$+0.006\pm0.004$&$-0.006\pm0.003$&$+0.035\pm0.003$\\
$0.8<|Z|$ (kpc)$\leq 1.2$ &       &$-0.012\pm0.008$&$-0.015\pm0.008$&$-0.017\pm0.006$&$-0.002\pm0.006$\\
$1.2<|Z|$ (kpc)$<2.0$     &       &$-0.022\pm0.018$&$+0.009\pm0.019$&$-0.037\pm0.012$&$-0.013\pm0.013$\\
\noalign{\smallskip}
\hline
\end{tabular}

\end{table*}

\begin{table*}[t]
\caption[]{Vertical abundance gradients measured in the RC RAVE sample for Fe, Mg, Al,
 Si, and Ti  expressed as dex kpc$^{-1}$ for four
ranges of $|Z|$. For comparison, in the last column, the vertical [Fe/H]
gradients of the mock sample are reported. Only stars in the range 7.5$\le$\R(kpc)$<$8.5 are
considered. Uncertainties of 68\% confidence are
obtained with the bootstrap method and represent the internal errors. 
}
\label{tab_RC_XH_grad_z}
\vskip 0.3cm
\centering
\begin{tabular}{l|ccccc|c}
\hline
\noalign{\smallskip}
        &  $\frac{d[Fe/H]}{dZ}$  & $\frac{d[Mg/H]}{dZ}$  &
$\frac{d[Al/H]}{dZ}$ & $\frac{d[Si/H]}{dZ}$ & $\frac{d[Ti/H]}{dZ}$ &
$\frac{d[Fe/H]}{dZ}$ (mock)\\
\noalign{\smallskip}
\hline
\noalign{\smallskip}
$0.0<|Z|$ (kpc)$\leq 0.4$ &$-0.050\pm0.027$&$+0.019\pm0.022$&$+0.045\pm0.030$&$-0.088\pm0.030$
&$+0.081\pm0.023$ & $-0.373\pm0.031$ \\
$0.4<|Z|$ (kpc)$\leq 0.8$ &$-0.087\pm0.030$&$+0.022\pm0.025$&$-0.027\pm0.034$&$-0.117\pm0.033$ &$+0.076\pm0.023$
& $-0.409\pm0.038$\\
$0.8<|Z|$ (kpc)$\leq 1.2$ &$-0.148\pm0.073$&$-0.031\pm0.061$&$-0.124\pm0.075$&$-0.197\pm0.078$ &$+0.010\pm0.055$
& $-0.536\pm0.085$\\
$1.2<|Z|$ (kpc)$<2.0$     &$-0.199\pm0.070$&$+0.041\pm0.096$&$+0.031\pm0.104$&$-0.140\pm0.096$ &$-0.086\pm0.068$
& $-0.338\pm0.083$\\
\noalign{\smallskip}
\hline
\end{tabular}

\end{table*}

\begin{table*}[t]
\caption[]{Radial abundance gradients for Fe, Mg, Al,
 Si, and Ti elements measured for the RC RAVE sample
divided in $\alpha$-poor (top) and $\alpha$-rich (bottom) sub-samples. The
gradients are expressed as dex kpc$^{-1}$ for four
ranges of $|Z|$. Uncertainties of 68\% confidence are
obtained with the bootstrap method and represent the internal errors. 
}
\label{tab_a_samp_grad_R}
\vskip 0.3cm
\centering
\begin{tabular}{l|ccccc}
\hline
\noalign{\smallskip}
        &  $\frac{d[Fe/H]}{dR}$  & $\frac{d[Mg/H]}{dR}$  & $\frac{d[Al/H]}{dR}$
& $\frac{d[Si/H]}{dR}$ & $\frac{d[Ti/H]}{dR}$\\
\noalign{\smallskip}
\hline
\noalign{\smallskip}
$\alpha$-poor sample & \\
$0.0<|Z|$ (kpc)$\leq 0.4$ &$-0.054\pm0.004$&$-0.037\pm0.004$&$-0.036\pm0.005$ & $-0.065\pm0.005$ & $+0.009\pm0.004$ \\
$0.4<|Z|$ (kpc)$\leq 0.8$ &$-0.045\pm0.004$&$-0.035\pm0.004$&$-0.041\pm0.005$ & $-0.050\pm0.005$ & $-0.008\pm0.003$\\
$0.8<|Z|$ (kpc)$\leq 1.2$ &$-0.026\pm0.008$&$-0.020\pm0.007$&$-0.031\pm0.010$ & $-0.039\pm0.009$ & $-0.018\pm0.007$\\
$1.2<|Z|$ (kpc)$<2.0$     &$+0.040\pm0.019$&$+0.030\pm0.016$&$+0.049\pm0.025$ & $+0.004\pm0.020$ & $+0.035\pm0.016$\\
\noalign{\smallskip}
\hline
\noalign{\smallskip}
$\alpha$-rich sample & \\
$0.0<|Z|$ (kpc)$\leq 0.4$ &$-0.010\pm0.018$&$-0.001\pm0.017$&$+0.016\pm0.023$ &$-0.003\pm0.019$ & $+0.021\pm0.019$ \\
$0.4<|Z|$ (kpc)$\leq 0.8$ &$+0.009\pm0.011$&$+0.001\pm0.011$&$+0.028\pm0.014$ &$-0.010\pm0.011$ & $+0.026\pm0.011$\\
$0.8<|Z|$ (kpc)$\leq 1.2$ &$-0.002\pm0.014$&$-0.018\pm0.014$&$-0.041\pm0.019$ & $-0.022\pm0.015$ & $-0.018\pm0.014$\\
$1.2<|Z|$ (kpc)$<2.0$     &$+0.047\pm0.028$&$+0.026\pm0.031$&$+0.085\pm0.043$ & $+0.009\pm0.026$ & $+0.019\pm0.039$\\
\noalign{\smallskip}
\hline

\end{tabular}
\end{table*}

\begin{table*}[t]
\caption[]{Vertical abundance gradients for Fe, Mg, Al,
 Si, and Ti elements measured for the RC RAVE sample
divided in $\alpha$-poor (top) and $\alpha$-rich (bottom) sub-samples.
Only stars in the range 7.5$\le$\R(kpc)$<$8.5 are
considered. The
gradients are expressed as dex kpc$^{-1}$ for four
ranges of $|Z|$. Uncertainties of 68\% confidence are
obtained with the bootstrap method and represent the internal errors. 
}
\label{tab_a_samp_grad_z}
\vskip 0.3cm
\centering
\begin{tabular}{l|ccccc}
\hline
\noalign{\smallskip}
        &  $\frac{d[Fe/H]}{dZ}$  & $\frac{d[Mg/H]}{dZ}$  &
$\frac{d[Al/H]}{dZ}$& $\frac{d[Si/H]}{dZ}$ & $\frac{d[Ti/H]}{dZ}$\\
\noalign{\smallskip}
\hline
\noalign{\smallskip}
$\alpha$-poor sample & \\
$0.0<|Z|$ (kpc)$\leq 0.4$ &$+0.004\pm0.025$&$+0.025\pm0.023$&$+0.071\pm0.029$ & $-0.048\pm0.030$ & $+0.095\pm0.023$ \\
$0.4<|Z|$ (kpc)$\leq 0.8$ &$-0.028\pm0.028$&$+0.009\pm0.026$&$+0.004\pm0.033$ & $-0.083\pm0.033$ & $+0.103\pm0.024$\\
$0.8<|Z|$ (kpc)$\leq 1.2$ &$-0.085\pm0.076$&$-0.131\pm0.064$&$-0.061\pm0.081$ & $-0.153\pm0.086$ & $+0.029\pm0.062$\\
$1.2<|Z|$ (kpc)$<2.0$     &$-0.141\pm0.090$&$-0.084\pm0.098$&$+0.108\pm0.143$ & $-0.066\pm0.120$ & $-0.121\pm0.085$\\
\noalign{\smallskip}
\hline
\noalign{\smallskip}
$\alpha$-rich sample &\\
$0.0<|Z|$ (kpc)$\leq 0.4$ &$-0.312\pm0.113$&$-0.189\pm0.114$&$-0.051\pm0.134$ &$-0.243\pm0.111$ & $+0.077\pm0.129$ \\
$0.4<|Z|$ (kpc)$\leq 0.8$ &$+0.015\pm0.092$&$-0.004\pm0.089$&$+0.004\pm0.115$ &$+0.061\pm0.094$ & $+0.062\pm0.104$\\
$0.8<|Z|$ (kpc)$\leq 1.2$ &$+0.141\pm0.120$&$+0.192\pm0.112$&$-0.077\pm0.174$ & $+0.071\pm0.144$ & $+0.133\pm0.125$\\
$1.2<|Z|$ (kpc)$<2.0$     &$-0.073\pm0.088$&$-0.016\pm0.108$&$+0.015\pm0.130$ & $-0.094\pm0.150$ & $+0.072\pm0.131$\\
\noalign{\smallskip}
\hline

\end{tabular}
\end{table*}

\subsection{Comparison with other observational works}
The negative radial gradients that become shallower as we move away from the Galactic plane
are in fair agreement (within the external errors, $\sim0.01$~dex kpc$^{-1}$)
with the results obtained with the DW sample in Paper~1,
although the values are not exactly comparable because for the DW sample 
we measured the radial gradients in the (\Rg,\Zmax) plane while here we considered the (\R,\z) plane
\footnote{When the (\Rg,\Zmax) plane is considered, the RC sample shows shallower
radial gradients but the trend is the same (i.e. the gradients
are shallower when the distance from the Galactic plane increases).}.\\
A similar conclusion can be drawn by considering the work of
Anders et al. \cite{anders} for their radial gradients mesured
on the ($R_\mathrm{med}$, \Zmax) plane with APOGEE stars ($R_\mathrm{med}$
is the median radius of the Galactic orbit). When the (\R,\z) plane is used,
the gradients of Anders et al. are in good agreement with the RAVE
results, but for $|z|<0.4$~kpc, where the gradient is steeper than the RAVE
gradient.\\

Hayden et al. \cite{hayden}, also employing APOGEE stars, found
negative inner- and outer-disc radial gradients, flatter for the former, steeper
for the latter. The gradients become flatter with \z\ and the break between the
inner- and outer-disc gradients progressively shifts to larger \R.
Although in the RAVE RC sample we do not see any break, our radial gradients
have values that seem to lie between the inner- and outer-disc gradients
found by Hayden et al., and they flatten with \z\ in the same manner as
Hayden's gradients.\\
Significant differences are found for the  $\alpha$-poor and $\alpha$-rich
samples. In the Hayden et al. sample both
$\alpha$-poor and $\alpha$-rich stars exhibit negative radial gradients,
while in our RAVE sample only the $\alpha$-poor stars exhibit negative radial gradients,
while the radial gradients of the RAVE $\alpha$-rich stars are consistent
with zero. The vertical gradients found by Hayden et al. are significantly
more negative  ($\sim-0.31$, $-0.21$ and $-0.26$~dex~kpc$^{-1}$ at $7<$\R(kpc)$<9$ 
for their whole sample, $\alpha$-poor sample, and $\alpha$-rich sample, respectively)
with respect to the RAVE RC sample ($\sim-0.11$, $-0.067$ and
$-0.021$~dex~kpc$^{-1}$ for the whole RAVE RC sample, $\alpha$-poor sample, and $\alpha$-rich
sample, respectively) in the $0<|Z|$(kpc)$<2$ range.\\

The radial gradients of the SEGUE sample studied by Cheng et al. \citealp{cheng_b}  
agree with our results with negative and zero radial gradients for 
the $\alpha$-poor and the $\alpha$-rich stars, respectively, although the
$\alpha$-poor stars of the SEGUE sample at $|Z|<0.5$~kpc are steeper than
the RAVE sample ($-0.1$ versus $-0.05$~dex~kpc$^{-1}$).
In the framework  of the thin-/thick-disc dichotomy, the zero radial gradients
exhibited by the $\alpha$-rich stars agree with other previous works in which the 
authors claimed no detectable radial gradient in the thick
disc (Co{\c s}kuno{\v g}lu et al. \citealp{coskunoglu} by using RAVE DR3 data; 
Ruchti et al. \citealp{ruchti}; Cheng et al. \citealp{cheng}).\\

In the SEGUE sample studied by Schlesinger et al. \cite{schlesinger}
the vertical gradient at the solar circle is $-0.243^{+0.039}_{-0.053}$~dex~kpc$^{-1}$
in the range $0.5\lesssim|Z|(\mbox{kpc})\lesssim1.7$.
For comparison purposes we measured the iron vertical
gradient of our RAVE RC sample in
the same $|Z|$ range obtaining $-0.159\pm0.016$~dex~kpc$^{-1}$,
which may be consistent with the Schlesinger et al. gradient inside a 2$\sigma$
error.
The SEGUE $\alpha$-poor stars show a positive vertical metallicity gradient
($+0.063^{+0.047}_{-0.032}$~dex~kpc$^{-1}$),
which might be consistent (inside a $2\sigma$ error) with our zero or
negative gradient
for $|Z|>0.4$~kpc. The vertical gradient of the SEGUE $\alpha$-rich stars
($+0.038^{+0.043}_{-0.037}$~dex~kpc$^{-1}$) is consistent with
our results. Unfortunately, they have no results for $|Z|<0.4$~kpc to be
compared
with our zero and negative gradients of the $\alpha$-poor and $\alpha$-rich
stars, respectively.\\

While the absence of a radial gradient in the thick disc seems
to be confirmed by different studies, the vertical
gradient is still debated.
Some studies found no significant vertical gradients
in the thick disc (Allende Prieto et al.
\citealp{allende_prieto}; Peng et al. \citealp{peng}), whereas others claim negative
vertical gradients (Bilir et al. \citealp{bilir} by using RAVE
DR3 data; Chen et al. \citealp{chen}; Ruchti et al. \citealp{ruchti}; Katz
et al. \citealp{katz}). Our work suggest that the vertical gradient of the 
thick disc may be consistent with zero. \\

\subsection{Comparison with models}
We stressed before that a detailed comparison of our observational
results with Galaxy models is beyond the scope of this
work. Nonetheless, in Sect.~\ref{sec_comp_mock} we compared our
results with the mock sample (which is a GALAXIA realization of the 
Besan\c con \citealp{robin} model) and here we would like very 
briefly to compare our results with a few chemical and
chemo-dynamical models found in the literature as a first check.\\
Great care must be taken in drawing conclusions when comparing results
from different sources. Observational data, as much as results from chemical
or chemo-dynamical models, may refer to samples of stars having different age
distribution, therefore referring to different evolutionary phases of the
Galaxy. Chemical simulations of the Galaxy showed that stars of different ages
can have significantly different radial gradients (Minchev et al. \citealp{minchev},
\citealp{minchev14b}; Sch\"onrich \& Binney \citealp{schoenrich}), therefore
a sample of stars covering a wide range of ages can exhibit gradients that are
a function of their age distribution. Our RAVE RC sample covers
presumably a wide range of ages but we do not know its exact age
distribution. Keeping this uncertainty in mind, we report here some results
from chemical and chemo-dynamical models.\\

In the $|Z|\lesssim0.4$~kpc range (Tab.~\ref{tab_RC_XH_grad_R}), 
our iron radial gradient is in
qualitative agreement with the predictions of chemical evolution models of
the Galaxy, which assume an inside-out formation process of the disc (Chiappini et
al. \citealp{chiappini2001}; Cescutti et al. \citealp{cescutti}; Gibson et al.
\citealp{gibson}). The predicted gradients of the
Cescutti et al. \cite{cescutti} model for Fe and Mg ($-0.052$~dex kpc$^{-1}$ and $-0.039$~dex
kpc$^{-1}$, respectively) are particularly close to our findings, 
whereas for Si ($-0.035$~dex kpc$^{-1}$) the gradient is too flat with respect to our
$-0.064$~dex kpc$^{-1}$ reported in Tab.~\ref{tab_RC_XH_grad_R}. 
The predicted Ti ($-0.032$~dex kpc$^{-1}$) does not agree with our
zero radial gradient found for this element. The metallicity gradients of
$\sim-0.05$~dex kpc$^{-1}$  predicted by the models of Minchev et al.
(\citealp{minchev}; \citealp{minchev14b}) for moderately young stars
(age$<$4~Gyr) can be in reasonable agreement with our RC sample. 
The model by Sch\"onrich \& Binney 
\cite{schoenrich} predicts a metallicity gradient ($-0.11$~dex kpc$^{-1}$)
too steep with respect to our results.

\subsection{On the constancy of the chemical gradients}
We estimate the gradients by fitting the data with linear laws,
which is convenient for a rough estimate but it may oversimplify the
reality. In the evolution of a galaxy, stochastic processes may play a
significant role (e.g. where and when an open cluster is going to burst, or
a disrupted satellite is accreted by the Galaxy), generating coarse
distributions of the stars in velocity and chemical space.
Irregularity in the distributions can be revealed only by accurate
measurements.
When accurate chemical abundances are available, the distributions of the
stars in the (\R,[X/H]) and (\z,[X/H]) planes are clearly not linear. 
Hayden et al. \cite{hayden} subdivide the radial metallicity
gradient observed with the APOGEE data into two linear laws; Balser et al. \cite{balser} found 
different radial gradients when they are measured at different Galactic
azimuth; by using SEGUE data  Schlesinger et al. \cite{schlesinger} show clear
wiggles in the iron abundance as a function of \z. 
Haywood et al. \cite{haywood} suggested that the negative radial gradients measured
may be the result of the superposition of an inner and an outer disc, which have
different average metallicities and $\alpha$-enhancement because they
experienced different star formation histories. In this scenario, the radial
metallicity gradient would not be linear but step-like. 
Thanks to the large amount of data expected from the coming big spectroscopic surveys
(such as Gaia, \citealp[Perryman et al.][]{perryman}; GALAH, \citealp[Zucker et
al.][]{zucker}; 4MOST, \citealp[de Jong et
al.][]{dejong}, among others) it will be possible to perform ``tomographies" of
the disc, and move from 1D to 3D in the analysis of the chemical gradients.

\section{Conclusions}\label{sec_conclusion}
In this work, we measured the radial and vertical chemical abundance gradients of the
Galactic disc in the range 4.5$<$\R(kpc)$<$9.5 
for the elements Mg, Al, Si, Ti, and Fe to provide
new constraints to the chemical Galactic models.
We selected a sample of 17\,950 giant stars (the RAVE RC sample) 
from the RAVE internal database on the basis of the gravity 
values in the range 1.7$<$\logg$<$2.8.
Similarly, we selected a corresponding sample of stars
from a mock RAVE sample created with the Galaxia code (Sharma et al.
\citealp{sharma}) and based on the Besan\c con model (Robin et al.
\citealp{robin}) to compare our data with a well-established
Galaxy model. The RAVE and the mock samples are the same samples
used in our previous paper (Boeche
et al. \citealp{boeche13}, Paper~1) in which we studied the radial gradients
by means of dwarf stars. We divided the RAVE and mock RC samples in
four consecutive $|Z|$ bins with boundaries at 0, 0.4, 0.8, 1.2, and 2~kpc to study
the radial gradients as a function of \z\ for the five elements under analysis. 
For a deeper analysis, we further subdivided the samples in
$\alpha$-poor ([Mg/Fe]$\lesssim$0.25~dex) and $\alpha$-rich
([Mg/Fe]$>$0.25~dex) stars.
Our major results can be summarized with the following points:
\begin{itemize}
\item the radial chemical gradients are negative and become progressively flatter with $|Z|$
\item the vertical chemical gradients are negative and become progressively steeper with
$|Z|$ (but seem to flat out at large $|Z|$) 
\item the $\alpha$-rich stars have radial chemical gradients consistent with zero
\item the vertical chemical gradients of the $\alpha$-poor stars are consistent 
with zero or with being slightly negative
\item the vertical chemical gradients of the $\alpha$-rich stars are consistent with zero but close 
to the Galactic plane, where they are negative.
\end{itemize}

These results are generally valid although there can be differences
depending on the element and the $|Z|$ interval considered. To fully address
the reasons of these differences a detailed comparison with Galactic
chemical models is needed.\\
Close to the Galactic disc the RAVE RC sample has a [Fe/H] radial gradient of
$-0.055$~dex kpc$^{-1}$, and it becomes flatter with $|Z|$. This is in good
agreement with previous works (Boeche et al. \citealp{boeche13}; Cheng et al. \citealp{cheng}; 
Hayden et al. \citealp{hayden}; Anders et al. \citealp{anders}).
This is also in agreement with the mock RC sample, which can reproduce the
radial gradient as well as its flattening with $|Z|$.
By studying the RAVE RC sample as a function of the $\alpha$-enhancement, 
we found that the $\alpha$-poor stars drive the radial gradients while the
$\alpha$-rich stars show radial gradients consistent with zero.
In the framework of the thin/thick disc duality, this suggests that the thick disc
may have no radial gradient. This supports the results of
previous studies (Co{\c s}kuno{\v g}lu et al. \citealp{coskunoglu}; 
Ruchti et al. \citealp{ruchti}; Cheng et al.
\citealp{cheng}, \citealp{cheng_b}).\\
The vertical chemical gradients exhibited by the RAVE RC sample are negative
($-0.11$~dex~kpc$^{-1}$) but shallower than the mock RC sample
($-0.40$~dex~kpc$^{-1}$). The steep gradient exhibited by the mock RC sample
originates from the difference in average metallicity (0.78~dex) between the
thin- and the thick-disc stars combined with their different scale-heights.
The same difference in average metallicity also explains the broader
metallicity distribution of the mock RC sample with respect to the RAVE RC
sample.
These discrepancies can be reduced by increasing the average [Fe/H] of 
the thick disc in the mock RC sample from the actual [Fe/H]=$-0.78$~dex
to $\sim-0.5$~dex, as suggested in Boeche et al. \cite{boeche13} 
and supported by other studies
(Soubiran et al. \citealp{soubiran03}, Ivezi{\'c} et al.
\citealp{ivezic}, Kordopatis et al. \citealp{kordopatis11}).
The proposed explanation of the negative vertical gradient as a
result of the overlap of two populations
with shallow or no vertical gradient (as the thin- and thick-disc in the
mock RC sample) finds support in the study of the vertical gradient of
the RAVE $\alpha$-poor and $\alpha$-rich samples.
The individual vertical gradients of these two samples are consistent with zero or 
being slightly negative. Therefore, similarly to the mock RC
sample and in the framework of the thin- thick-disc duality, 
the negative vertical gradient observed in the RAVE RC 
sample could be mainly caused by the superposition of two
populations, one having high metallicity and small scale-height and
the other having low metallicity and large scale-height.\\
The vertical iron gradient observed in the RAVE RC sample seems slightly
flatter than found by Schlesinger et al. \cite{schlesinger} and
definitely flatter than found by Hayden et al. \cite{hayden}.
We found that the $\alpha$-rich sample has a significant negative [Fe/H] vertical
gradient at $|Z|<0.4$. This is consistent with the work by Minchev et al.
\cite{minchev14} who, by using RAVE data, found an unexpected small vertical 
velocity dispersion for stars with [Mg/Fe]$\gtrsim$0.4~dex, suggesting
radial migration of stars from the inner disc as result
of massive mergers in the early history of the Galaxy.\\

Our current knowledge of the Galactic disc tells us that it cannot be
homogeneous and that a unique radial (or vertical) chemical gradient value cannot be valid for the
whole disc. 
Our work and other studies (Schlesinger et al. \citealp{schlesinger};
Hayden et al. \citealp{hayden}; Balser et al. \citealp{balser}), show that the
chemical gradients can vary vertically and radially as well as in
Galactic azimuth, and that the linear laws used to describe the change of 
chemical abundances through the Galactic disc seem to be an oversimplification. 
Future large spectroscopic surveys like the Gaia-ESO survey
\cite[Gilmore et al.][]{gilmore12}, the GALactic Archaeology with HERMES
survey (GALAH) \cite[Zucker et al.][]{zucker}, 
the Apache Point Observatory Galactic Evolution Experiment
survey (APOGEE) \cite[Majewski et al.][]{majewski}, the Large sky Area
Multi-Object fiber Spectroscopic Telescope (LAMOST) \cite[Zhao et
al.][]{zhao}, and Gaia \cite[Perryman et al.][]{perryman}
will be able to bring new
information about the spatial and chemical distribution of the Galactic disc,
and will permit us to perform a ``tomography" of the disc, disentangling the
spatial chemical distribution due to the Galactic disc formation from the
expected inhomogeneities due to moving groups, disrupted open clusters, and
stellar streams due to merging Galactic satellites.

\begin{acknowledgements}
We thank the anonymous referee for the constructive comments
that helped to improve the paper. 
B.C. thanks J. Rybizki the useful discussions
during the preparation of this work. 
We acknowledge funding from Sonderforschungsbereich SFB 881 ``The Milky Way
System" (subproject A5) of the German Research Foundation (DFG).  Funding
for RAVE has been provided by the Australian Astronomical Observatory; the
Leibniz-Institut fuer Astrophysik Potsdam (AIP); the Australian National
University; the Australian Research Council; the French National Research
Agency; the German Research Foundation (SPP 1177 and SFB 881); the European
Research Council (ERC-StG 240271 Galactica); the Istituto Nazionale di
Astrofisica at Padova; The Johns Hopkins University; the National Science
Foundation of the USA (AST-0908326); the W.  M.  Keck foundation; the
Macquarie University; the Netherlands Research School for Astronomy; the
Natural Sciences and Engineering Research Council of Canada; the Slovenian
Research Agency; the Swiss National Science Foundation; the Science \&
Technology Facilities Council of the UK; Opticon; Strasbourg Observatory;
and the Universities of Groningen, Heidelberg and Sydney.  The RAVE web site
is at http://www.rave-survey.org.

\end{acknowledgements}


\begin{thebibliography}{}

\bibitem[2003]{abadi} Abadi, M.~G., Navarro, 
J.~F., Steinmetz, M., \& Eke, V.~R.\ 2003, \apj, 597, 21 
%
\bibitem[2006]{allende_prieto} Allende Prieto, 
C., Beers, T.~C., Wilhelm, R., et al.\ 2006, \apj, 636, 804 
%
\bibitem[2013]{anders} Anders, F., Chiappini, 
C., Santiago, B.~X., et al.\ 2013, arXiv:1311.4549 
%
\bibitem[2011]{balser} Balser, D.~S., Rood, 
R.~T., Bania, T.~M., \& Anderson, L.~D.\ 2011, \apj, 738, 27
%
\bibitem[2006]{belokurov} Belokurov, V., 
Zucker, D.~B., Evans, N.~W., et al.\ 2006, \apjl, 642, L137 
%
\bibitem[1994]{bertelli} Bertelli, G., Bressan, A., Chiosi, C.,
Fagotto, F., \& Nasi, E.\ 1994, \aaps, 106, 275 
%
\bibitem[2012]{bijaoui} Bijaoui, A., Recio-Blanco, A., de Laverny, P., et
al.\ 2012, StMet, 9, 55
%
\bibitem[2012]{bilir} Bilir, S., Karaali, S., 
Ak, S., et al.\ 2012, \mnras, 421, 3362 
%
%
\bibitem[2014]{binney} Binney, J., Burnett, B., 
Kordopatis, G., et al.\ 2014, \mnras, 437, 351
%
\bibitem[2011]{boeche11} Boeche, C., Siebert, A., Williams, M., et al., 2011,
\aj, 142, 193
%
\bibitem[2013]{boeche13} Boeche, C, Siebert, A., Piffl, T., et al.\
2013, \aap, 559, A59
%
\bibitem[2012]{bovy} Bovy, J., Rix, H.-W., 
\& Hogg, D.~W.\ 2012, \apj, 751, 131 
%
\bibitem[2011]{brunetti11} Brunetti, M., Chiappini, C., Pfenniger, D. 2011, A\&A 534, 75
%
\bibitem[1985]{carlberg85} Carlberg, R.~G., 
\& Sellwood, J.~A.\ 1985, \apj, 292, 79 
%
\bibitem[2007]{cescutti} Cescutti, G., Matteucci, F., Fran{\c c}ois, P., \&
Chiappini, C.\ 2007, \aap, 462, 943
%
\bibitem[2011]{chen} Chen, Y.~Q., Zhao, G., 
Carrell, K., \& Zhao, J.~K.\ 2011, \aj, 142, 184 
%
\bibitem[2012a]{cheng} Cheng, J.~Y., Rockosi, 
C.~M., Morrison, H.~L., et al.\ 2012, \apj, 746, 149 
%
\bibitem[2012b]{cheng_b} Cheng, J.~Y., Rockosi, 
C.~M., Morrison, H.~L., et al.\ 2012, \apj, 752, 51
%
\bibitem[1997]{chiappini} Chiappini, C., 
Matteucci, F., \& Gratton, R.\ 1997, \apj, 477, 765 
%
\bibitem[2001]{chiappini2001} Chiappini, C., 
Matteucci, F., \& Romano, D.\ 2001, \apj, 554, 1044 
%
\bibitem[2012]{coskunoglu} Co{\c 
s}kuno{\v g}lu, B., Ak, S., Bilir, S., et al.\ 2012, \mnras, 419, 2844
%
\bibitem[1998]{dehnen} Dehnen, W., Binney, J.\ 1998, \mnras, 294, 429
%
\bibitem[2012]{dejong} de Jong, R.~S., 
Bellido-Tirado, O., Chiappini, C., et al.\ 2012, \procspie, 8446,
%
\bibitem[2002]{freeman} Freeman, K., \&
Bland-Hawthorn, J.\ 2002, \araa, 40, 487 
%
\bibitem[2002]{friel} Friel, E.~D., Janes, 
K.~A., Tavarez, M., et al.\ 2002, \aj, 124, 2693
%
\bibitem[2013]{gibson} Gibson, B.~K., Pilkington, K., Brook, C.~B.,
Stinson, G.~S., \& Bailin, J.\ 2013, \aap, 554, A47 
%
\bibitem[1983]{gilmore} Gilmore, G., \& Reid, N.\ 1983, \mnras,
202, 1025 
%
\bibitem[2012]{gilmore12} Gilmore, G., Randich, 
S., Asplund, M., et al.\ 2012, The Messenger, 147, 25 
%
%
\bibitem[2014]{hayden} Hayden, M.~R., Holtzman, 
J.~A., Bovy, J., et al.\ 2014, \aj, 147, 116
%
\bibitem[2013]{haywood} Haywood, M., Di Matteo, P., Lehnert, M.~D.,
Katz, D., \& G{\'o}mez, A.\ 2013, \aap, 560, A109 
%
\bibitem[2008]{ivezic} Ivezi{\'c}, {\v Z}., 
Sesar, B., Juri{\'c}, M., et al.\ 2008, \apj, 684, 287 
%
\bibitem[2011]{katz} Katz, D., Soubiran, C., Cayrel, R., et al.\
2011, \aap, 525, A90 
%
\bibitem[2011]{kobayashi} Kobayashi, C., \& Nakasato, N.\
2011, \apj, 729, 16 
%
\bibitem[2011]{kordopatis11} Kordopatis, G., Recio-Blanco, A., de
Laverny, P., et al.\ 2011, \aap, 535, A107 
%
\bibitem[2013]{kordopatisDR4} Kordopatis, G., 
Gilmore, G., Steinmetz, M., et al.\ 2013, \aj, 146, 134 
%
\bibitem[2002]{kroupa} Kroupa, P.\ 2002, \mnras, 330, 
707 
%
\bibitem[2000]{hog} H{\o}g, E., Fabricius, C., Makarov, V.~V., et al.\ 2000,
 \aap, 355, L27 
%
\bibitem[2011]{lee} Lee, Y.~S., Beers, T.~C., 
An, D., et al.\ 2011, \apj, 738, 187
%
\bibitem[2006]{luck} Luck, R.~E., Kovtyukh, 
V.~V., \& Andrievsky, S.~M.\ 2006, \aj, 132, 902 
%
\bibitem[2010]{majewski} Majewski, S.~R., 
Wilson, J.~C., Hearty, F., Schiavon, R.~R., 
\& Skrutskie, M.~F.\ 2010, IAU Symposium, 265, 480 
%
\bibitem[2008]{marigo} Marigo, P., Girardi, L., Bressan, A., et
al.\ 2008, \aap, 482, 883
%
\bibitem[2012]{matijevic} Matijevi{\v c}, 
G., Zwitter, T., Bienaym{\'e}, O., et al.\ 2012, \apjs, 200, 14 
%
\bibitem[2010]{mf10} Minchev, I., \& Famaey, B.\  2010, \apj, 722, 112 
%
\bibitem[2011]{minchev11a} Minchev, I., Famaey, B., Combes, F., et al.\ 2011, \aap, 527, A147
%
%
\bibitem[2013]{minchev} Minchev, I., Chiappini, C., \& Martig, M.\
2013, \aap, 558, A9 
%
\bibitem[2014]{minchev14} Minchev, I., Chiappini, 
C., Martig, M., et al.\ 2014, \apjl, 781, L20 
%
\bibitem[2014b]{minchev14b} Minchev, I., Chiappini, 
C., \& Martig, M.\ 2014, arXiv:1401.5796 
%
\bibitem[2001]{montes} Montes, D., 
L{\'o}pez-Santiago, J., G{\'a}lvez, M.~C., et al.\ 2001, \mnras, 328, 45 
%
\bibitem[2004]{nordstrom04} Nordstr\"om, B., Mayor, M., Andersen, J., et al., 2004, A\&A 418, 989
%
\bibitem[1993]{pasquali} Pasquali, A., Perinotto, M.\ 1993, A\&A 280, 581
%
\bibitem[2013]{peng} Peng, X., Du, C., Wu, Z., 
Ma, J., \& Zhou, X.\ 2013, \mnras, 434, 3165
%
\bibitem[2001]{perryman} Perryman, M.~A.~C., de Boer, K.~S., Gilmore, G., et
al. 2001, A\&A 369, 339
%
\bibitem[1993]{quinn} Quinn, P.~J., Hernquist, 
L., \& Fullagar, D.~P.\ 1993, \apj, 403, 74 
%
\bibitem[2006]{recio-blanco} Recio-Blanco, A., 
Bijaoui, A., \& de Laverny, P.\ 2006, \mnras, 370, 141 
%
\bibitem[2003]{robin} Robin, A.~C., Reyl{\'e}, C., Derri{\`e}re,
S., \& Picaud, S.\ 2003, \aap, 409, 523 
%
\bibitem[2008]{roeser2008} R{\"o}ser, S., Schilbach, E., Schwan, H., et
al.\ 2008, \aap, 488, 401 
%
\bibitem[2010]{roeser2010} Roeser, S., Demleitner, 
M., \& Schilbach, E.\ 2010, \aj, 139, 2440 
%
\bibitem[2008]{roskar08} Ro{\v s}kar, R., 
Debattista, V.~P., Quinn, T.~R., Stinson, G.~S., 
\& Wadsley, J.\ 2008, \apjl, 684, L79
%
\bibitem[2011]{ruchti} Ruchti, G.~R., 
Fulbright, J.~P., Wyse, R.~F.~G., et al.\ 2011, \apj, 737, 9 
%
\bibitem[2009]{sales} Sales, L.~V., Helmi, A., 
Abadi, M.~G., et al.\ 2009, \mnras, 400, L61 
%
\bibitem[2014]{schlesinger} Schlesinger, K., J., Johnson, J., A., Rockosi , 
C., M., et al., 2014, ApJ accepted,  arXiv:1405.6724 
%
\bibitem[2009]{schoenrich} Sch{\"o}nrich, R., \& Binney, J.\
2009, \mnras, 396, 203 
%
\bibitem[2002]{sellwood02} Sellwood, J.~A., \& Binney, J.~J.\
2002, \mnras, 336, 785 
%
\bibitem[2011]{sharma} Sharma, S., 
Bland-Hawthorn, J., Johnston, K.~V., \& Binney, J.\ 2011, \apj, 730, 3 
%
\bibitem[2011]{siebert} Siebert, A., Williams, M.~E.~K., Siviero, A., et
al.\ 2011, \aj, 141, 187
%
\bibitem[2003]{soubiran03} Soubiran, C., Bienaym\'e, O., Siebert, A.\ 2003,
A\&A, 398, 141
%
\bibitem[2006]{rave} Steinmetz, M., Zwitter, T., Siebert, A., et al.\ 2006,
\aj, 132, 1645
%
\bibitem[1995]{teuben} Teuben, P.J., The Stellar Dynamics Toolbox NEMO, in: 
Astronomical Data Analysis Software and Systems IV, 
ed. R. Shaw, H.E. Payne and J.J.E. Hayes. (1995),
PASP Conf Series 77, 398
%
\bibitem[2008]{villalobos} Villalobos A., Helmi A., 2008, \mnras, 391, 1806
%
\bibitem[2004]{zacharias} Zacharias, N., Urban, S.~E., Zacharias, M.~I., 
et al.\ 2004, \aj, 127, 3043
%
%
%
\bibitem[2009]{yanny} Yanny, B., Rockosi, C., Newberg, H.~J., et al.\ 2009, \aj, 137, 4377 
%
\bibitem[2012]{zhao} Zhao, G., Zhao, Y.-H., Chu, Y.-Q., Jing, Y.-P., \&
Deng, L.-C.\ 2012, Research in Astronomy and Astrophysics, 12, 723
%
\bibitem[2012]{zucker} Zucker, D.~B., de Silva, 
G., Freeman, K., Bland-Hawthorn, J., 
\& Hermes Team 2012, Galactic Archaeology: Near-Field Cosmology and the
Formation of the Milky Way, 458, 421 
%
\end{thebibliography}
\end{document}